\documentclass[manuscript]{aastex}   

\shorttitle{The Serpens Cloud in CO J=$2-1$ and \tco~J=$2-1$ Emission}
\shortauthors{Burleigh et al.}

\newcommand{\kms}{km~s$^{-1}$}

\newcommand{\msun}{M$_{\odot}$}
\newcommand{\degree}{$^{\circ}$}
\newcommand{\tco}{$^{13}$CO}        
\newcommand{\tb}{$T_{b}$}
\newcommand{\ntwoh}{N$_{2}$H$^{+}$}

\newcommand{\sigv}{$\sigma_{v}$}
\newcommand{\tempb}{$T_{b,max}$}
\newcommand{\tkin}{$T_{kin}$}

\newcommand{\Av}{$A_{V}$}
\newcommand{\vd}{$V_{drift}$}
\newcommand{\msigv}{$<\sigma_{v, CO}>$}

\newcommand{\squig}{$\sim \,$}
\newcommand{\cmvol}{cm$^{-3}$ \,}

\def\av#1{\langle#1\rangle}	      

\begin{document}


\title{
The Arizona Radio Observatory\\CO Mapping Survey of Galactic Molecular Clouds:\\
III. The Serpens Cloud in CO J=$2-1$ and \tco~J=$2-1$ Emission}

\author{Kaylan J. Burleigh\altaffilmark{1,2}}
\email{kaylanb@berkeley.edu}

\author{John H. Bieging\altaffilmark{2}, Alisha Chromey\altaffilmark{2,3}, Craig Kulesa\altaffilmark{2}, \and William L. Peters\altaffilmark{2} }

\altaffiltext{1}{Astronomy Department, University of California, Berkeley CA 94720 USA}
\altaffiltext{2}{Steward Observatory, The University of Arizona, Tucson, AZ 85721, USA}
\altaffiltext{3}{Dept. of Physics and Astronomy, Iowa State University, Ames IA 50011, USA}

\received{22 June 2013}
\accepted{05 November 2013}


\begin{abstract}
We mapped $^{12}$CO and \tco~J = $2-1$ emission over 1.04 deg$^2$ of the Serpens molecular cloud with $38\arcsec$ spatial and 0.3 \kms~ spectral resolution using the Arizona Radio Observatory Heinrich Hertz Submillimeter telescope. Our maps resolve kinematic properties for the entire Serpens cloud. We also compare our velocity moment maps with known positions of Young Stellar Objects (YSOs) and 1.1 mm continuum emission. We find that $^{12}$CO is self-absorbed and \tco~ is optically thick in the Serpens core. Outside of the Serpens core, gas appears in filamentary structures having LSR velocities which are blue-shifted by up to 2 \kms~ relative to the 8 \kms~ systemic velocity of the Serpens cloud. We show that the known Class I, Flat, and Class II YSOs in the Serpens core most likely formed at the same spatial location and have since drifted apart. The spatial and velocity structure of the $^{12}$CO line ratios implies that a detailed 3-dimensional radiative transfer model of the cloud will be necessary for full interpretation of our spectral data. The ``starless cores" region of the cloud is likely to be the next site of star formation in Serpens.
\end{abstract}


\keywords{ISM: clouds - ISM: individual objects (Serpens cloud) - ISM: kinematics and dynamics - ISM: molecules}






\section{Introduction}

\subsection{The Serpens Molecular Cloud}

The Serpens cloud is a low mass star-forming cloud in the Gould Belt. The cloud is known for its high star formation rate (SFR) and high surface density of Young Stellar Objects (YSOs) (Eiroa et al. 2008). A 10 deg$^2$ optical extinction (A$_V$) map made by Cambresy (1999) originally defined the Serpens cloud. More recent studies treat the Serpens cloud as two much smaller ($\sim$ 1.0 deg$^{2}$) regions: Serpens Main (centered on R.A. $18^{h}29^{m}00^{s}$,  Dec. $+00^{\circ}30'00''$ (J2000)) and Serpens South (centered on R.A. $18^{h}30^{m}00^{s}$, Dec. $-2^{\circ}02'00''$ (J2000)) (Enoch et al. 2007; Harvey et al. 2007a; Gutermuth et al. 2008; Eiroa et al. 2008; Bontemps et al. 2010). Serpens Main is known mainly for its northernmost region, the Serpens core, which has the highest YSO surface density in the Cambresy (1999) A$_V$ map (Eiroa et al. 2008). Serpens South is part of the Aquila rift molecular complex. It was first studied in detail by Gutermuth et al. (2008), and has now been mapped at 70 - 500 $\mu$m as part of the {\it Herschel} Gould Belt Survey (Andre et al. 2010; Bontemps et al. 2010). Serpens Main is the focus of this study (see  Fig. \ref{fig:mapped_reg}).

The total molecular mass of the Serpens core is uncertain by at least a factor of 5. Some studies estimate $\sim$250-300 \msun~ (McMullin et al. 2000; Olmi \& Testi 2002) while others find $\sim$1450 \msun~(White et al. 1995). These two results were based on C$^{18}$O J=1-0 and C$^{18}$O J=$2-1$ lines, respectively, so the large discrepancy may be due to the different gas properties traced by each C$^{18}$O rotational transition (Eiroa et al. 2008). The distance to Serpens assumed by these studies may also be too low (see Section 1.2). Clearly, the gas mass must be measured with better accuracy to determine the efficiency and history of star formation in the Serpens cloud.

\subsection{Distance to Serpens Main}

The best estimate for the distance to Serpens Main is $415 \pm 25$ pc, measured from VLBI trigonometric parallax of the YSO, EC 95 (Dzib et al. 2010). EC 95 is located at the center of the Serpens core and is therefore almost certainly a member, so we adopt the Dzib et al. (2010) 415 pc distance. This is almost twice the previously accepted value of $230 \pm 20$ pc (Eiroa et al. 2008), so care must be used in comparing physical properties derived using the lower distance with our results in this paper.

\subsection{Our Study}

\begin{figure}
\epsscale{0.7}
\plotone{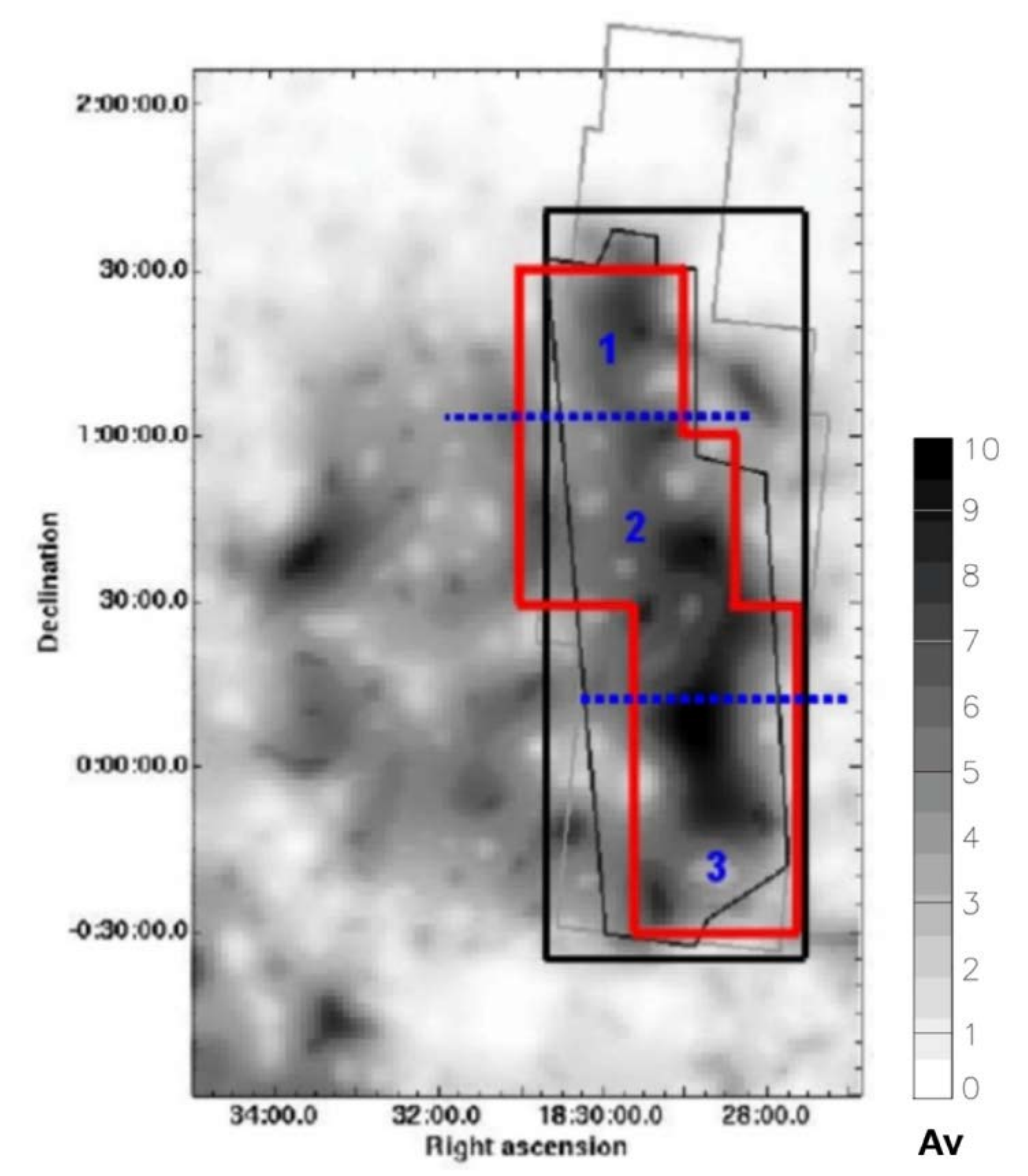}
\caption{ The Serpens Main molecular cloud (figure adapted from Enoch et al. 2007). Gray scale map is a sub-section of the Cambresy et al. (1999) $A_{V}$ map of Serpens, scale bar gives $A_{V}$ in mag. The figure shows our $^{12}$CO and \tco~ mapped region (red polygon), c2d IRAC (thin black polygon) and MIPS (gray polygon) observations of Serpens Main (Evans et al. 2003), and Bolocam observations of 1.1 mm emission (thick black polygon) (Enoch et al. 2007). Regions labeled 1 -- 3 and separated by dotted lines (blue) correspond to the Serpens core, Serpens G3-G6, and VV Serpentis regions, respectively. The Cambresy (1999) map shows \Av~ $\leq$ 10 mag, but the c2d derived \Av~ map shows \Av~$\gtrsim$ 25 mag (Enoch et al. 2007).
	\label{fig:mapped_reg}}
\end{figure}

We mapped about 1.04 deg$^{2}$ of Serpens Main in the CO and \tco~J = $2-1$ emission lines. Our study complements existing survey data, from the {\it Spitzer} c2d Legacy (Evans et al. 2003; Harvey et al. 2006; Harvey et al. 2007) and Bolocam 1.1 mm Continuum (Enoch et al. 2007) Surveys. Other molecular tracer data exist for Serpens Main, such as \ntwoh~ (Testi et al. 2000), but these data are almost always limited to a sub-region of Serpens Main (e.g., the Serpens core). Fig. \ref{fig:mapped_reg} shows our mapped region (red polygon), the {\it Spitzer} c2d regions with MIPS (gray polygon) and IRAC (thin-black polygon), and the Bolocam 1.1 mm region (thick black polygon) overlaid on the Cambresy (1999) \Av~ map of Serpens Main. The Cambresy (1999) map suggests \Av~ $\leq$ 10 mag in Serpens Main, but the more recent \Av~ map derived from c2d {\it Spitzer} data (Enoch et al. 2007) shows \Av~$\gtrsim$ 25 mag. Continuum 1.1 mm emission reveals the locations of the coldest and densest dust. \ntwoh~ traces the highest density star forming gas (Testi et al. 2000).
 
From the {\it Spitzer} c2d data, Harvey et al. (2007) presented a ``high-confidence set" of 235 YSOs associated with Serpens Main. Most YSOs have masses $< 1.4$ \msun~ (Eiroa et al. 2008), but there is at least one (VV Serpentis) with mass $\approx 4 - 5$ \msun~ (Ripepi et al. 2007; Dzib et al. 2010). Enoch et al. (2007) identified 35 sub-mm sources in Serpens Main from the Bolocam 1.1 mm Survey. We will compare the positions of these YSOs and sub-mm sources with our CO and \tco~ data (see Section 5).

We divide Serpens Main into three sub-regions with the following naming scheme: the {\it Serpens core} (Eiroa et al. 2008), {\it Serpens G3-G6} (Cohen \& Kuhi 1979), and {\it VV Serpentis} (Chavarria et al. 1988). These sub-regions are labeled {\it 1, 2,} and {\it 3}, respectively, in Fig. \ref{fig:mapped_reg}. Other names previously used for the Serpens core include Serpens dark, Serpens North, and Cluster A (Harvey et al. 2007; Eiroa et al. 2008; Bontemps et al. 2010); Cluster B for Serpens G3-G6 (Harvey et al. 2007; Enoch et al. 2007); and Cluster C for VV Serpentis (Harvey et al. 2007; Eiroa et al. 2008).

This study is a continuation of a molecular cloud mapping project with the Arizona Radio Observatory. Previous papers in this series mapped the W51 region in CO J = $2-1$ and \tco~ J = $2-1$  (Bieging, Peters, \& Kang 2010) and the W3 region in CO J = $2-1$ and J = $3-2$ and \tco~ J = $2-1$ (Bieging \& Peters 2011). For further details about the Serpens Cloud, we refer the reader to: Harvey et al. (2007); Enoch et al. (2007); Gutermuth et al. (2008); Eiroa et al. (2008); and Bontemps et al. (2010).
 
\subsection{Goal}

The goal of our study was to map the Serpens core, Serpens G3-G6, and VV Serpentis regions with high resolution in the J=$2-1$ rotational lines of $^{12}$C$^{16}$O and $^{13}$C$^{16}$O (hereafter CO and \tco~respectively). CO emission will be determined principally by  cloud temperature and global turbulence. In contrast, \tco~will generally be more useful as a measure of the column density. 

Section 2 describes our observations and data reduction techniques. In Section 3, we show our final brightness temperature image cubes for CO and \tco~J=2-1, and present associated velocity moment maps. Section 4 gives the website where our calibrated CO and \tco~ image cubes can be downloaded. In Section 5, we overlay the Harvey et al. (2007) YSOs, Enoch et al. (2007) sub-mm sources, and Testi et. al (2000) \ntwoh cores on our CO and \tco~ maps. Section 6 is Discussion, and section 7 is our summary.

\section{Observations and Data Reduction}

The observations were made between 2008 November and 2010 June with the Heinrich Hertz Submillimeter Telescope (HHT) on Mt. Graham, AZ, at an elevation of 3200 m. The HHT has a 10-m diameter paraboloidal dish and observes in the frequency range from 210 to 500 GHz. 
Using prototype ALMA Band 6 sideband separating mixers (courtesy of the National Radio Astronomy Observatory) in a dual polarization receiver on the HHT (Lauria et al. 2006), we were able to take CO (230.5 GHz) and \tco~(220.4 GHz), J=$2-1$, spectra simultaneously through the same telescope optics. 

We used a filterbank spectrometer to record both the CO and \tco~ emission  simultaneously.  Each of the two J=$2-1$ lines was measured with 256 filters of 0.25~MHz bandwidth, giving a total spectral coverage of $\sim41$ \kms~ at a resolution of 0.33 \kms.  
CO and \tco~ J = $2-1$ emission in our mapped region of Serpens spans a velocity range from $-5$ to 20 \kms~ (LSR). Our spectra extend from $-11$ to 30 \kms~(LSR) so we detect all gas in the cloud. 

Observations were made by the ``on-the-fly" (OTF) method, where the telescope was scanned in a boustrophedonic pattern in right ascension at a scan rate of 10$\arcsec$/sec, with spectra sampled every 0.1 sec.  In subsequent processing, the data were smoothed by 4 samples or 0.4 sec, corresponding to a telescope motion of 4$\arcsec$ per spectrum.  The beam size at the CO J=$2-1$ observing frequency is 32$\arcsec$ (FWHM) so this choice of spatial sampling does not cause any significant loss of angular resolution in the scan direction.  A raster of lines with 10$\arcsec$ spacing in declination or $\sim0.3$ of the beam FWHM was observed in this way.  The telescope resolution is well-sampled in both coordinates with this choice of OTF mapping parameters.  The field to be mapped was divided into 39 subfields, each $10\arcmin \times 10\arcmin$ in size, plus a small overlap.  Each subfield was observed at least once in each of the two CO isotopes.  Total scanning time per subfield was therefore 2.0 hours, plus $\sim$30\% overhead for telescope slewing and measurement of reference spectra after every other row.  Subfields were re-observed in whole or part if weather conditions resulted in excessively high noise levels.  

The telescope pointing accuracy was checked at the start of each observing period by measuring the CO J=$2-1$ emission from the red giant star V1111 Oph, which has strong CO emission lines from its circumstellar molecular envelope.  This object is a good pointing calibrator because the stellar emission is compact and centered on the well-known position of the star, which is only about 10\degree~from our Serpens field.  A 5-point cross pattern centered on the star was measured and the integrated line intensities at the 5 positions were fitted to determine the pointing corrections in azimuth and elevation.  These corrections were generally $<5\arcsec$.  The pointing was checked and corrected about every 2 hours throughout each observing run, so the pointing errors during the OTF mapping should be $<5\arcsec$.

The intensity scale was determined with the standard method of Kutner \& Ulich (1981) by comparing an ambient temperature load with on-sky spectra near the target field, to give the spectral line intensity on the $T_A^*$~scale.  The sideband rejections for the mixers were measured each time the receiver was tuned.  We observed a standard position, W51D, to place all the OTF data on a consistent main-beam brightness temperature scale, in the manner described in Bieging et al. (2010).  

All of the OTF data were processed in the {\it CLASS} software package of the Grenoble Astrophysics Group.  A linear baseline was removed from each spectrum and the data for each spectral channel were convolved to a square grid in R.A. and Dec., with a 10$\arcsec$ grid spacing in both coordinates.  The gridding algorithm uses a circular gaussian weighting function with a FWHM of 0.3$\times$(beam FWHM), to convolve the OTF-sampled spectra onto the regular grid points.  The gridding thereby increased the effective angular resolution by a factor of $(1+0.3^2)^{1/2} = 1.044$.

The resulting data cubes were transferred to the {\it MIRIAD} software format (Sault, Teuben, \& Wright 1995) for subsequent processing and analysis.  To facilitate the comparison of CO and \tco ~lines (for example, to calculate line ratios at each pixel), we used the {\it MIRIAD} task {\it regrid} to resample the data cubes on the velocity axis by 3rd order interpolation.  We chose to sample velocities at 0.15 \kms ~intervals from $-11$ to $+30$ \kms ~(LSR)\footnote{~All velocities in this paper refer to the Local Standard of Rest.}, so that the original velocity resolution was oversampled by a factor of 2.2.  

To match the angular resolutions of the CO and \tco~ J=$2-1$ maps, which differ by $\sim4\%$ because of the difference in line rest frequencies, we convolved the images with gaussians chosen to give identical resolutions for the two isotopologues.  The effective resolution of the final maps was 38$\arcsec$ (FWHM) for both the CO and \tco ~images.  This additional spatial smoothing also reduced the surface brightness noise level, at only a small cost in angular resolution.  

\begin{figure}
\epsscale{1.}
\plottwo{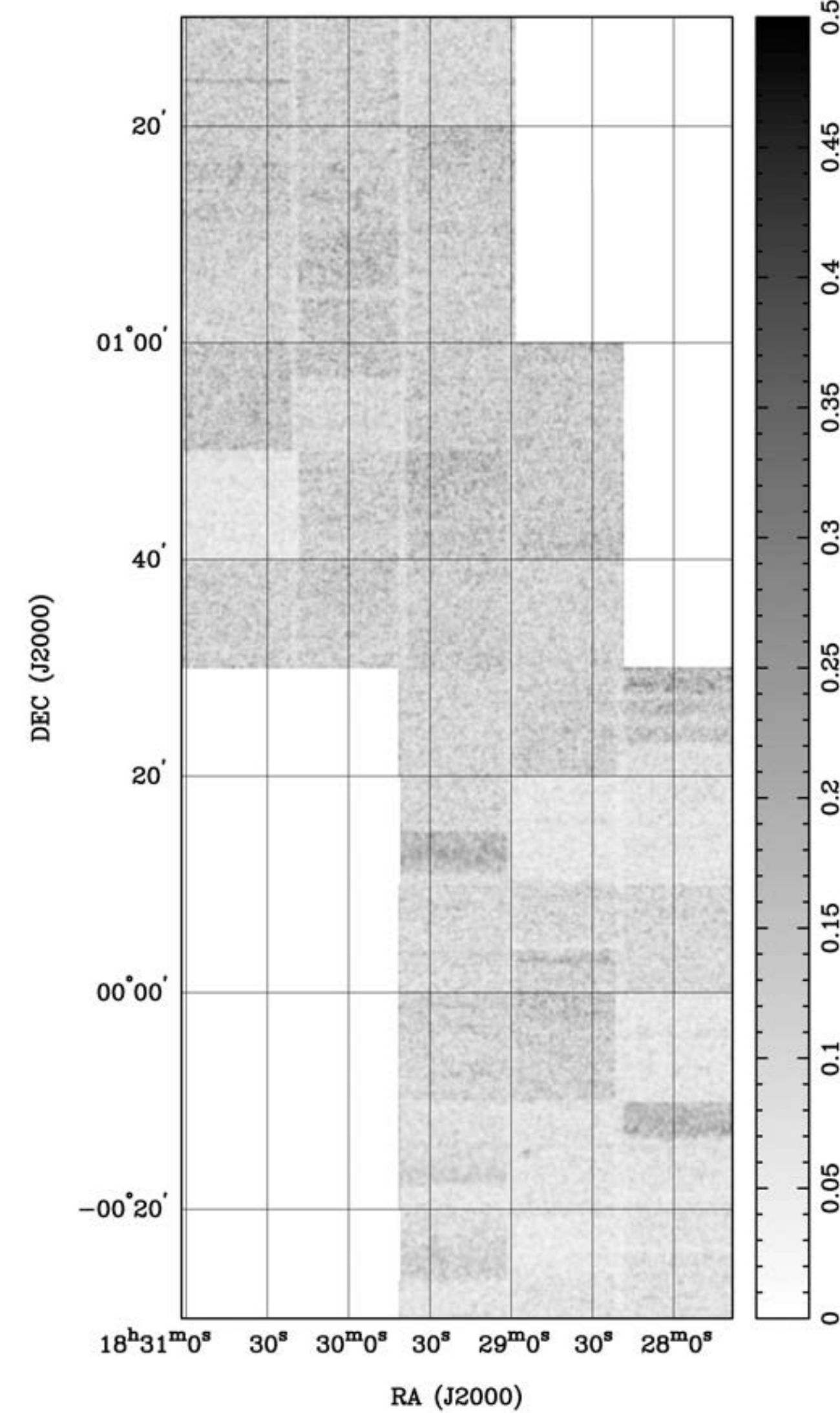}{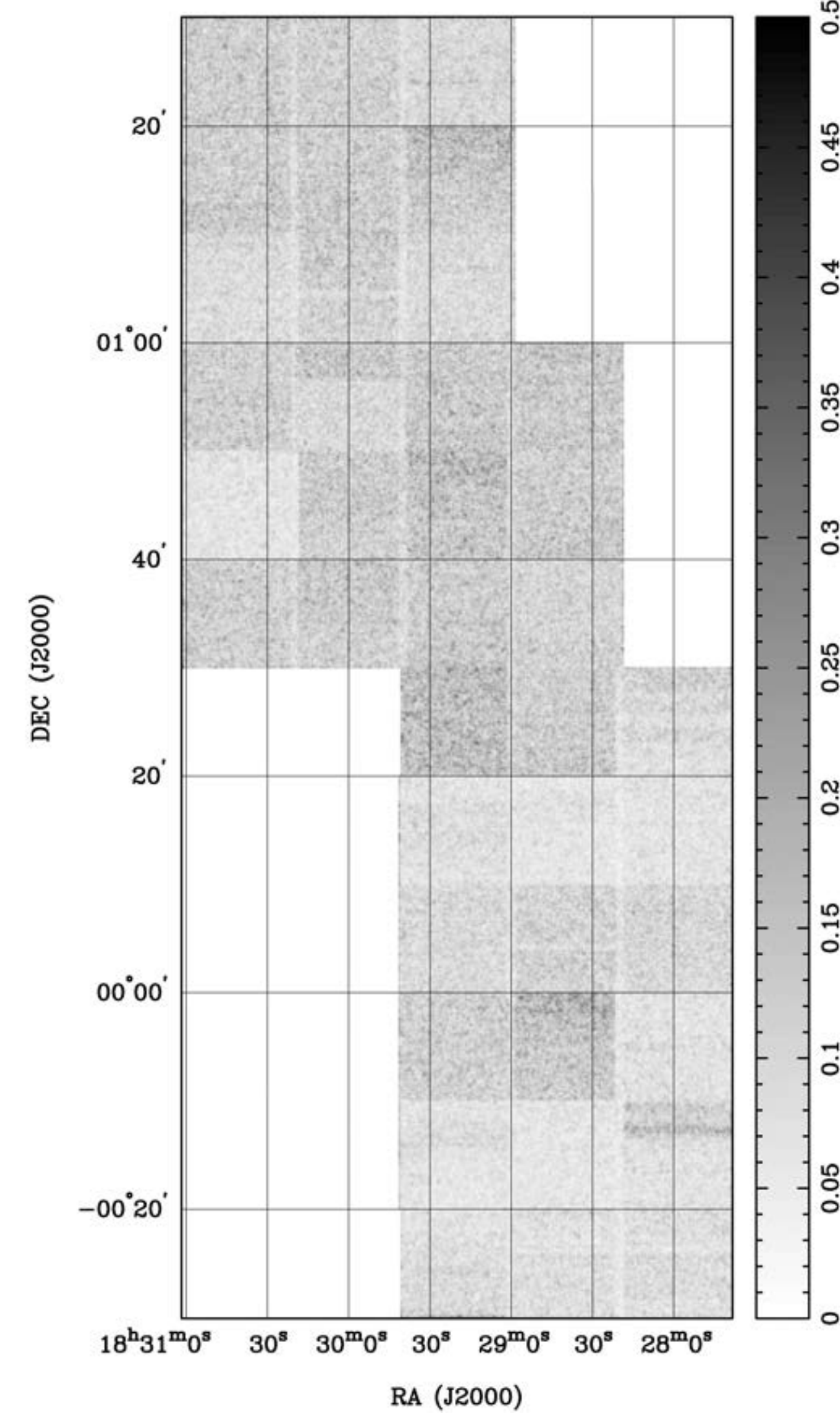}
\caption{Grayscale maps showing RMS noise level at each pixel of the CO J=$2-1$ {\it (left)} and $^{13}$CO J=$2-1$ {\it (right)} emission maps. Units of intensity wedges are main beam brightness temperature in Kelvins.    
	\label{fig:rms_map}}
\end{figure}

We calculated the RMS noise in an emission-free velocity range for each map pixel.  Fig. \ref{fig:rms_map} shows the resulting distribution of the noise.  The CO J=$2-1$ image {\it (left)} has relatively uniform noise over the whole field, with a mean value of $\av{RMS} = 0.106$~K (main beam brightness) per pixel and per velocity channel, with variations of $<35\%$ about this value over most of the map.  For the entire \tco~image {\it (right)} the mean $\av{RMS} = 0.105$~K with comparable levels of variation due to differences in weather (i.e., sky noise) and source elevation when the individual subfields were observed.

\section{Results}

\subsection{CO and \tco~ spectra and channel maps}

\begin{figure}
\epsscale{1.1}
\plottwo{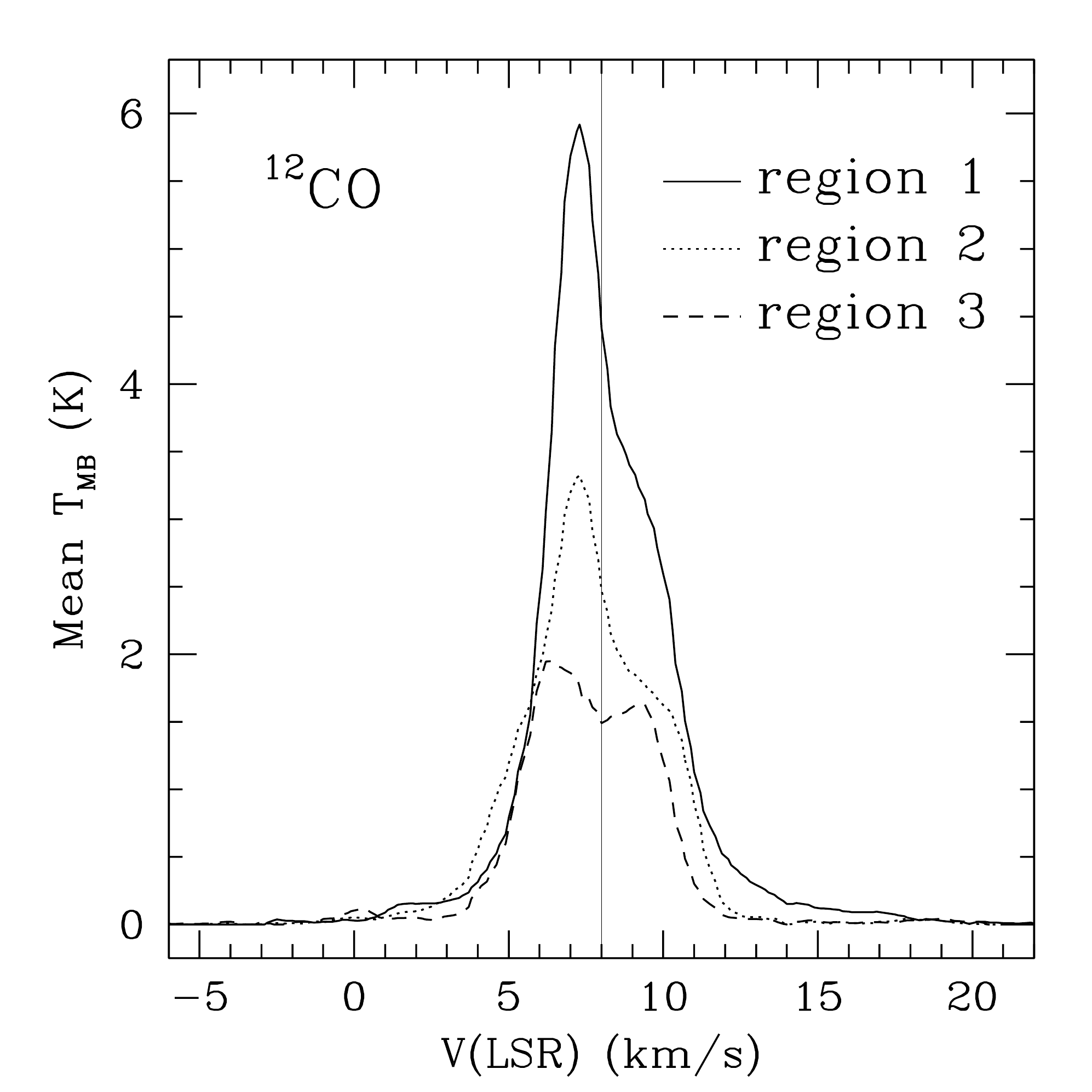}{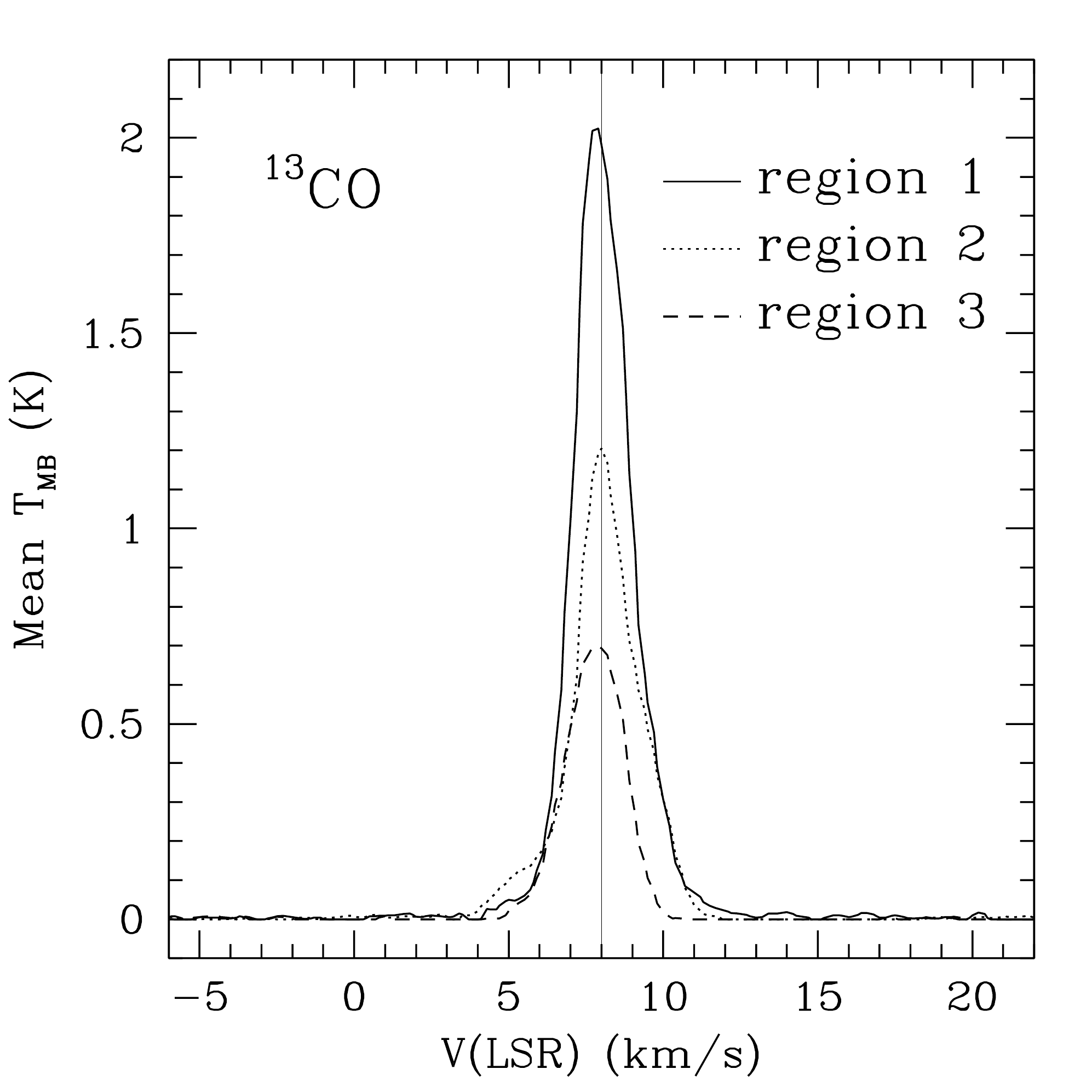}
\caption{Mean brightness temperature spectra averaged over the regions 1, 2, \& 3 indicated in Fig. \ref{fig:mapped_reg}: {\it (left)}, CO J=$2-1$ and {\it (right)}, \tco~J=$2-1$.  Vertical line marks the nominal systemic velocity of 8 \kms.   
	\label{fig:mean_spectra}}
\end{figure}

In Fig. \ref{fig:mean_spectra} we show the spatially averaged CO and \tco~spectra for the Serpens core (region 1), Serpens G3-G6 (region 2), and VV Serpentis (region 3) in Fig. \ref{fig:mapped_reg}.  The radial velocity of the emission ranges from about $-1$ to 18 \kms~ for CO and from 2 to 13 \kms~ for \tco.  The peak brightness temperature for \tco~ occurs at 8 \kms~ in all three regions, consistent with  previous studies (Duarte-Cabral et al. 2010; Graves et al. 2010) which concluded that this is the  velocity of the bulk of the gas in Serpens.  We adopt 8 \kms~ as the systemic velocity of Serpens (indicated by the vertical line in Fig. \ref{fig:mean_spectra}).  The \tco~ spectra for all three regions are nearly Gaussian in shape.  The CO profiles, in contrast, are markedly asymmetric and show a dip or inflection at 8 \kms, the velocity of the \tco~ peak emission.  This comparison strongly suggests that the CO lines are self-absorbed over at least some of the brightest emission regions, possibly as a result of colder gas which is optically thick in CO and is located on the near side of a warmer central part of the cloud.  

Fig. \ref{fig:co_channel} shows representative velocity slices from our CO image cube, averaging over 0.6 \kms~ and stepping by 1.2 \kms.  At velocities of about 4.8 and 10.8 \kms~ the emission appears elongated and filamentary.

Fig. \ref{fig:13co_channel} shows a similar set of velocity slices from our \tco~ image cube, averaging over 0.6~\kms, at the same velocities as in Fig. \ref{fig:co_channel} for the brightest CO emission (4.4 to 10.4 \kms).  Outside this velocity range, \tco~is mostly below our detection limit.  The \tco~ J=$2-1$ emission, which is generally optically thin, should be a good tracer of molecular column density in the cloud. 

\subsection{Maps of velocity moments}

Fig. \ref{fig:peak_bt} shows maps of maximum brightness temperature ($T_{b,max}$) in CO and \tco~, over the full range of emission, -11 to 30 \kms~ for CO and -5 to 20 \kms~ for \tco. The CO \tempb~ distribution is very clumpy with temperatures of 5 to 10 K in Serpens G3-G6 and parts of VV Serpentis, and temperatures of 15 to 25 K in the Serpens core. When CO is not self-absorbed or depleted, CO \tempb~ approaches the value of the kinetic temperature of the gas (\tkin). In contrast, \tco~ \tempb~ is much more uniform.

We present integrated CO and \tco~ brightness temperature (\tb) maps in Fig. \ref{fig:integrated_bt}, displayed in units of K \kms. We only consider velocities associated with Serpens by summing over -1 to 18 \kms~ for CO, and 2 to 13 \kms~ for \tco. We use a 3 RMS cutoff to ensure that detected emission is real. \tco~ is generally optically thin and its spectrum well-behaved, so the \tco~ integrated \tb~ map is nearly identical to the \tco~ $T_{b,max}$ map. \tco~ integrated \tb~ should trace the highest column density gas, as long as \tco~ is optically thin. Compared to \tempb~ for CO, CO integrated \tb~ retains its clumpy structure in the Serpens G3-G6 region, but is smoothly distributed otherwise. 

We show centroid velocity maps (first-moment) for CO and \tco~ in Fig. \ref{fig:vel_centroid}, considering the same range of velocities as we used to make the integrated \tb~ maps. Note that even where CO is self-absorbed, the first moment is the mean velocity of the entire line profile and not (necessarily) the velocity of the highest peak bracketing the self-absorption. The color palette extends $\pm$ 3 \kms~ on either side of 8 \kms~, the systemic LSR velocity of Serpens. Red colors are redshifted with respect to the systemic velocity. The centroid velocity of CO in the Serpens core and \tco~ in all regions is mostly at the systemic velocity of Serpens and spatially uniform. This uniformity contrasts with the centroid velocity of CO southward of the Serpens core, which is not spatially uniform but appears in filamentary structures with LSR velocities $\pm$ 3 \kms~ of the systemic velocity of Serpens. This difference in apparent spatial distribution is most likely due to a change in the velocity structure and/or the opacity of the colder absorbing CO outside the Serpens core. There are two kinematically distinct areas in the Serpens core, seen as red and blue components in CO centroid velocity separated by about 3 \kms. It has been suggested that these are two interacting sub-clouds (Testi et al. 2000; Eiroa et al. 2008; Duarte-Cabral et al. 2010). 

Fig. \ref{fig:vel_width} shows velocity dispersion ($\sigma_{v}$) (second-moment) maps for CO and \tco, in units of \kms. We apply a 3 RMS cutoff $\approx 0.35$ K for both isotopes. The velocity range for computing the CO $\sigma_{v}$ map is -1 to +16 \kms~ (which avoids the high velocity gas component at R.A. $18^h29^m12^s$, Dec. $+0\degr18\arcmin$), and 4 to 12 \kms~ for the \tco~ $\sigma_{v}$ map. The maximum CO $\sigma_{v}$ is 4.2 \kms, and that of \tco~ is 2.0 \kms, while the average CO \sigv~ is 2 \kms, and that of \tco~ is 0.8 \kms. Large $\sigma_{v}$ reveals extended emission-line wings, which can be caused by large scale gas motions or turbulence due to protostellar outflows. The \tco~ map is relatively noisy (i.e., pixelated) for small ($\lesssim 0.8$ ~\kms) \sigv, which occurs away from regions of high column density. Unlike \tco~ \tempb, integrated \tb, and large CO $\sigma_{v}$ ($\gtrsim 2.5$ \kms), large \tco~ $\sigma_{v}$ ($\gtrsim 1.5$ \kms) occurs in small (0.01 x 0.01 deg$^{2}$) regions. There is one such region in the Serpens core and three to five in Serpens G3-G6.

Note that we exclude a $0.2^{\circ} \times 0.2^{\circ}~$ high velocity (18 - 20 \kms) gas component (R.A. $18^h29^m12^s$, Dec. $+0\degr18\arcmin$) from our moment maps. We assume that this feature is not associated with Serpens Main.

\subsection{CO/\tco~ Ratio Maps}

Following Bieging et al. (2011), Fig. \ref{fig:ratio_map} shows the ratio ($R$) of CO to \tco~ line intensities (\tb), averaged over 0.6 \kms~ and spaced 0.6 \kms, as a function of velocity. For an assumed $^{12}$CO/$^{13}$CO abundance ratio of 50, the color palette indicates where \tco~ reaches optical depth $\tau_{^{13}CO} \geq 1$ for $1 \le R \le 1.6$ (gray regions), $\tau_{^{13}CO} < 1$ for $R > 1.6$ (colored regions), and CO is self-absorbed for $R \le 1$ (white regions). There is a spatial gradient in optically thick \tco~ across Serpens, appearing in VV Serpentis at low velocities and shifting north-easterly until \tco~ is optically thick in the Serpens core at higher velocities. CO is self-absorbed and \tco~ is optically thick in the Serpens core. The spatial and velocity structure of the CO line ratios imply that a detailed 3-dimensional radiative transfer model of the cloud will be necessary for proper interpretation of these data.

\section{Online-only Data}

The calibrated brightness temperature image cubes of CO and \tco~J=2-1 can be downloaded as FITS files from the online version of this paper through The Astrophysical Journal Supplement Series.

\section{Comparison with other observations}

\subsection{Distribution of YSOs and dense N$_2$H$^+$ cores}

The evolutionary stage of a YSO (after protostar formation but before reaching the main sequence) can be inferred from its observed SED (slope of log$_{10} \lambda$ F$_{\lambda}$ vs. log$_{10} \lambda$, $\lambda > 2~ \micron$) (Lada 1987; Andre \& Montmerle 1994; Greene et al. 1994) and/or comparing its SED to theoretical models (Whitney et al. 2003; Robitaille et al. 2006). These evolutionary stages are (in order of youngest to oldest): Class I, Flat, Class II, and Class III (Lada 1987; Andre \& Montmerle 1994; Greene et al. 1994). Harvey et al. (2007) classified their high-confidence set of 235 Serpens YSOs into the above four stages, 41 Class I (17 \%), 25 Flat (11 \%), 130 Class II (55 \%), and 39 Class III (17 \%). Of the 15 YSOs with the ``coldest SEDs" (70 to 24 $\micron$ flux ratio $> 8$, Harvey et al. 2007), 10 are Class I. These 10 (which we refer to as cold Class I) should be the youngest Class I YSOs as they are the most embedded or obscured. Note that Harvey et al. (2007) state that there are 39 Class I and 132 Class II YSOs, while the online supplementary table from Harvey et al. (2007) shows 41 Class I and 130 Class II. We use the data from the Harvey et al. (2007) online supplementary table. 

In Fig. \ref{fig:yso_olay}, we compare our \tco~ integrated intensity map with locations of the Harvey et al. (2007) YSOs: ``cold" Class I (magenta $\bigcirc$'s), all other Class I (magenta $+$'s), Flat (yellow $+$'s), Class II (white $\bigcirc$'s), and Class III (white $+$'s). The distribution of Class I and Flat YSOs follows regions of large ($\gtrsim 8$ K \kms) \tco~ integrated line intensity. There are three Class I YSOs (near R.A. 18$^h$28$^m$45$^s$, Dec. $+0\degr53\arcmin$) that do not coincide with large \tco~ integrated line intensity (see Section 5.3). In the Serpens core, all ``cold" Class I YSOs are in locations of peak ($\gtrsim 20$ K \kms) \tco~ integrated line intensity. Class II and III YSOs appear more widely distributed across Serpens Main.

Fig. \ref{fig:serpens_core} shows the \tco~ velocity centroid (1st moment) for the Serpens core computed over line center (6 to 10.5 \kms) in red-green-blue palette, with positions of N$_{2}$H$^{+}$ J=$1-0$ cores (cyan $\bigcirc$'s with diameter 1') from Testi et al. (2000) and YSOs (symbols as in Fig. $~\ref{fig:yso_olay}$) from Harvey et al. 2007. The N$_{2}$H$^{+}$ cores and Class I and Flat YSOs form a cluster that is strikingly elongated, with a major axis nearly parallel to the boundary separating the red and blue-shifted gas in the Serpens core.  The four \ntwoh~ cores, which should mark the coldest, densest pre-stellar locations in the cloud, also coincide closely with the youngest YSOs.  Class II and III YSOs are less clustered and more symmetrically distributed about this axis. This distribution raises the question of whether the younger (Class I and Flat) or older (Class II and III) YSOs, or both, formed along this boundary.

To examine these possibilities, we define two reference positions for Serpens Core YSOs: the unweighted centroid R.A., Dec. of cold Class I YSOs ({\it cold centroid}, R.A. $18^h29^m53\fs5$, Dec. $+1\degr15\arcmin16\farcs2$) and that of flat YSOs ({\it flat centroid}, R.A. $18^h29^m56\fs8$, Dec. $+1\degr13\arcmin27\farcs1$). Fig. \ref{fig:serpens_core} shows these two positions as black $\oplus$`s. The flat centroid lies within a tight grouping of flat YSOs \squig2\arcmin~ southeast of the cold centroid. 

We note that these two centroid positions lie near the centers of the two ``sub-clusters" described by Duarte-Cabral et al. (2010), based on their observations of the very optically thin C$^{17}$O and C$^{18}$O isotopologues.  Their ``NW sub-cluster" encompasses the position of our {\it cold centroid}, consistent with a very young age for the YSOs contained in this sub-cluster.  In contrast, Duarte-Cabral et al. (2010) find that the ``SE sub-cluster" displays greater kinematic complexity and a wider distribution of YSO ages.  This SE sub-cluster contains our {\it flat centroid} position, consistent with the observed clustering of the flat-SED sources having somewhat greater ages than the cold Class I sources.

In Figures \ref{fig:yso_hist_sep} and \ref{fig:yso_hist_pa}, we plot histograms of YSO projected separation and position angle (PA) relative to the cold (solid black) and flat (dashed blue) centroids, separated by YSO class. (Here ``Class I" includes ``cold Class I" YSOs.) The sample contains only YSOs within the Serpens core.  This restriction gives 23 Class I, 12 Flat, 23 Class II, and 7 Class III objects.

The histograms for the youngest classes (I and Flat) in Fig. 13 show that these YSOs cluster within 0.5 pc of either centroid position, with small offsets ($\sim$0.2 pc) as expected from the definitions of the 2 centroids.  The distribution of PAs for the youngest classes (Fig. 14) clearly shows that the clustering is oriented along a NW-SE ($315\degr - 135\degr$)~ axis, an orientation which is consistent with the location of the ``sub-clusters" discussed by Duarte-Cabral et al. (2010).  In contrast, the older YSOs, Classes II and III, show a wider distribution relative to either of the centroids (Fig. 13), with the Class III YSOs lying almost entirely at separations $>0.5$~pc, while the Class II YSOs are broadly peaked out to $\sim$1 pc.  Both the Class II and III objects are uniformly distributed in PA, with no preference for the NW-SE axis of the younger YSO classes.

The spatial distrobutions of the various YSO classes suggest that the most recent star formation in the Serpens core, i.e., within the past few$\times 10^5$~years, has occurred almost entirely within a volume of space about 1 pc in length and elongated in a NW-SE direction as projected on the sky.  The wider distribution of the Class III objects could be a result of their having formed over a more extended volume in an earlier episode of star formation, as suggested by Duarte-Cabral et al. (2010).  The Class II YSOs, however, show a distribution that is clustered about the reference centroid positions but less tightly than the Class I or Flat YSOs.  The progression from a more concentrated to a more dispersed distribution with age could also result from a diffusion of YSOs outward away from the Serpens core over the past few million years, due to random motions acquired from the internal velocity dispersion of the parent gas clouds.

\subsection{Maps of other CO isotopologues and transitions}

Davis et al. (1999) presented CO J=2-1 maps of the Serpens core region ($\sim 15\arcmin \times 12\arcmin$) with 23\arcsec~ resolution, and found evidence in the line wings (14 to 22 km s$^{-1}$ and $-4$ to 4 km s$^{-1}$) for a ``burst of outflows", produced by the cluster of YSOs in the core.  Our data have higher sensitivity but lower angular resolution (38\arcsec) than Davis et al. (1999);  a detailed comparison with their line wing maps shows excellent agreement for both the red- and blue-shifted components, including the broad red-shifted wing associated with the Herbig-Haro object HH106 (located $\sim 6\arcmin$ west of the core cluster).

More recently, Graves et al. (2010) have mapped the J=3-2 lines of CO, $^{13}$CO, and C$^{18}$O over the Serpens core with the HARP heterodyne receiver array on the JCMT, with effective resolutions of 17\arcsec~ to 20\arcsec. Their CO and $^{13}$CO integrated intensity maps (their Fig. 2) agree well with ours (Fig. 7), allowing for the different transitions and angular resolution.  At the position of HH106 (18$^h$29$^m$18$^s$, $+1\degr 14\arcmin 10\arcsec$), the CO J=2-1 line shows the same broad red-shifted wing as their J=3-2 line (their Fig. 5), but the lower excitation J=2-1 transition has a more pronounced narrow self-absorption feature at 8 km s$^{-1}$, the systemic velocity of the cloud.  Our line ratio channel maps (Fig. 7) show that near the systemic velocity, much of the Serpens core cloud is optically thick in $^{13}$CO J=2-1, and the CO line is likely to be self-absorbed.  Graves et al. (2010) also find from the J=3-2 line ratios that the $^{13}$CO line is optically thick with $\tau_{\rm 13}$ having values up to $\sim$7 toward the densest submm continuum cores.

Duarte-Cabral et al. (2010) presented maps of the C$^{17}$O and C$^{18}$O J=2-1 and 1-0 transitions made with the IRAM 30 m telescope, covering only the central 3\arcmin$\times$3.5\arcmin~ of the Serpens core.  These optically thin transitions should be less distorted by opacity effects than the $^{13}$CO or CO lines, revealing more accurately the distribution and kinematics of the highest column density regions.  Duarte-Cabral et al. (2010) found that the sub-clusters have different kinematic properties.  The NW center has a single, nearly Gaussian line profile, while the SE center has 2 blended components.  They interpret the profiles as a superposition of 2 clouds with different velocities and argue that these are colliding in the area containing the SE sub-cluster.  

In a second paper, Duarte-Cabral et al. (2011) show that a smoothed particle hydrodynamics (SPH) 3D simulation of two colliding gas cylinders, one rotated $\sim$ 45\degree~ relative to the other, can reproduce many of the observational properties of the Serpens core.  Especially the onset of star formation in the zone where the clouds collide. Their SPH simulation includes gas hydrodynamics, self-gravity, turbulence, and sink particles. 

Given the large differences in optical depths of the C$^{17}$O and C$^{18}$O lines compared with the CO and $^{13}$CO transitions we present here, it is difficult to make direct comparisons.  We note however, that Duarte-Cabral et al. (2010) find a velocity gradient in the C$^{18}$O J=1-0 map with velocity increasing from the SE to NW sub-clusters, that is consistent with that in our $^{13}$CO velocity centroid map (Fig. 12).  Our data show a gradient over a much larger region ($\sim$10\arcmin) than that covered by Duarte-Cabral et al. (2010, 2011).  Their SPH simulation may therefore underestimate the physical size of the colliding clouds.

\subsection{1.1 mm and submillimeter continuum emission}

In Fig. \ref{fig:bolocam_olay}, we overlay contours of 1.1 mm continuum emission (Enoch et al. 2007) on our \tco~ integrated \tb~ map (convolved to 90\arcsec~ resolution to match that of the published Bolocam map). Contour levels are at 5, 10, 20, 40, 80, and 160 times the RMS noise $\approx 11$ mJy beam$^{-1}$. The 1.1 mm emission coincides with high \tco~ integrated \tb, except for a filamentary feature in Serpens G3-G6 centered at R.A. $18^h28^m45^s$, Dec. $+0\degr53\arcmin$ called the ``starless cores region" (Enoch et al. 2007).  Three Class I YSOs and two Class II YSOs are located in the starless cores region (see Fig. \ref{fig:yso_olay}). We detect only a faint ($\lesssim 5$ K \kms) local maximum in \tco~ integrated \tb~ at R.A. $18^h28^m43^s$, Dec. $+0\degr54\arcmin$.

Submm maps of the Serpens core have been presented by Davis et al. (1999), made with the SCUBA bolometer array at wavelengths of 850~\micron~ and 450~\micron.  These images have considerably better resolution (14\arcsec~ and 8\arcsec~ respectively) but like the Bolocam 1.1 mm map, also suffer from spatial filtering so that mainly the bright compact sources are detected.  Four of these (SMM 1, 5, 9, and 10) lie within the NW sub-cluster and lie on the secondary maximum of the integrated $^{13}$CO intensity (Fig. 7, right) $\sim$3\arcmin~ NW of the brightest $^{13}$CO peak.  The other 6 submm compact sources are associated with the brightest peak in Fig. 7 (right), and with the SE sub-cluster of Duarte-Cabral et al. (2010).  This positional agreement implies that, as expected, the current most active star formation coincides with the region of greatest gas column density.

\subsection{{\it Herschel} Data}

Serpens Main has yet to be extensively studied using data from the {\it Herschel} Gould Belt Survey, which spans far-IR to sub-mm wavelengths. Such data could reveal even younger prestellar cores (Class 0) in Serpens Main. For example, Goicoechea et al. (2012) presented a submm spectrum for a single Serpens core Class 0 YSO.  Hundreds of Class 0 YSOs have been identified in the nearby Aquila rift molecular complex using {\it Herschel} data (Andre et al. 2010; Bontemps et al. 2010). 

\section{Discussion}

The spatial distribution of CO, as seen in the peak and integrated brightness temperature maps (Figs. 6 \& 7) as well as the individual channel maps (Fig. 4) gives the impression of a highly disturbed gas cloud.  There are shell-like structures and holes especially in the Serpens Core region.  Elsewhere the CO has a flocculent appearance with many small clumps ($\sim$0.1\degr) all having approximately the same peak brightness temperature.  Overall, there is a gradient of increasing peak and integrated brightness temperatures from south to north over the 2\degr~ declination extent of our maps.  Although CO emission is detected in every pixel, there are large contrasts in some regions.  The Serpens Core, at the northern end of the cloud, stands out sharply against a lower $T_B$ background, as a result of the high concentration of YSOs which energize the gas by radiation heating of the associated dust, and by injecting kinetic energy via numerous bipolar outflows.

The Class I and Flat YSOs in the Serpens core form a cluster that is nearly parallel to the intersection of the red and blue-shifted gas components and is strikingly linear-shaped (see Fig. \ref{fig:serpens_core}), similar to the corresponding distribution of sub-mm cores (Enoch et al. 2007; Duarte-Cabral et al. 2010). The velocity gradient might result from a collective action of outflows in the center of the Serpens core, and the linear-shaped cluster might indicate a preferred orientation for proto-stellar jets.  Duarte-Cabral et al. (2011) show that a cloud-cloud collision also explains the velocity gradient and linear-shaped cluster.   However, their SPH simulation should be treated with caution until the result is independently verified using an adaptive mesh refinement (AMR) code, such as Athena or FLASH, to calculate accurately the effects of shocks in the collision. 

From the observed \tco~velocity dispersion (Fig. \ref{fig:vel_width}) and present YSO distribution (Fig. \ref{fig:yso_hist_sep}), we find that Class I, flat, and Class II YSOs in the Serpens core had sufficient drift velocities and lifetimes to reach their present locations, even if they all formed in the same limited volume. Most Class I and flat YSOs are within \squig 0.5 projected pc of both the flat and cold centroids, while most Class II YSOs are within \squig 1.0 projected pc (see Fig. \ref{fig:yso_hist_sep}). Assuming a mean inclination angle of 45\degree~ to the line of sight, Class I and flat, and Class II YSOs would need to drift \squig 0.7 pc and 1.4 pc, respectively. Class II YSOs have lifetimes \squig~ 2 Myr, which sets Class I and flat YSO lifetimes at \squig 0.5 and 1.0 Myr, respectively (Evans et al. 2009). Assuming drift velocity (\vd) is constant since YSO formation, Class I, flat, and Class II YSOs would need minimum \vd~ \squig 2.0, 1.0, and 1.0 \kms, respectively. 

Because CO traces bulk gas motion, we can assume that \vd~ \squig~ \msigv~ (i.e., protostars inherit the velocity dispersion of accreted gas), where \msigv~ is the spatial average of $\sigma_{v,CO}$. We calculate \msigv~ over R.A. $18^h29^m20^s$ -- $18^h30^m23^s$ and Dec. $+1\degr09\arcmin16\arcsec$ -- $+1\degr21\arcmin16\arcsec$, the smallest region containing the YSO sample from Figs. 13 -- 14. \msigv~ = 2.0 \kms~ and RMS $\sigma_{v,CO} = 0.6$ \kms, giving \vd~ $= 2 \pm 0.6$ \kms. \vd~ exceeds or equals what is necessary for nearly all Class I, flat, and Class II YSOs to have formed in a small volume at either the flat or cold centroid positions.


Since we do not detect CO and \tco~ in the ``starless cores" region, CO is most likely depleted there. Average CO \tempb~ from Fig. \ref{fig:peak_bt} is \squig 10 - 20 K, so \tkin \squig 10 - 20 K (assuming no CO self-absorption or depletion). At 10 - 20 K CO becomes depleted for number densities $\gtrsim 3 \times 10^{4}$ \cmvol~ (Goldsmith 2001). The lowest (inferred) gas number density that Bolocam can detect is \squig $2 \times 10^{4}$ \cmvol (Enoch et al. 2007). CO is unlikely to be depleted in the Serpens core and Serpens G3-G6 regions because the temperature of the surrounding gas is too high, so CO will not freeze out onto dust grains, despite sufficiently high number densities. There are in fact five YSOs in the ``starless cores" region (i.e., not truly starless), but heating may be insufficient (CO and \tco~ peak T$_{B}$ is about 3 and 1.5 K, respectively) to prevent CO from freezing out onto gains.

\section{Summary}

We mapped the J=$2-1$ rotational lines of CO and \tco~ over 3900 square arcmin of the Serpens core, Serpens G3-G6, and VV Serpentis regions of the Serpens Main cloud (1.04 deg$^{2}$ in total) with high spatial (38\arcsec) and spectral (0.3 \kms) resolution. Our final data are calibrated \tb~ image cubes for CO and \tco~ (available online in FITS format). The {\it Spitzer} c2d Legacy Survey (Evans et al. 2003; Harvey et al. 2007) and the Bolocam 1.1 mm Continuum Survey (Enoch et al. 2007) are the only other studies that have mapped the entirety of the Serpens Main molecular cloud. These latter two surveys were limited to measuring dust temperature and column density; our CO and \tco~ data measure gas kinematics as well as physical properties.

At the systemic velocity of Serpens Main (8 \kms), CO is self-absorbed and \tco~ is optically thick in the Serpens core. The gas traced by CO in Serpens G3-G6 and VV Serpentis appears in filamentary structures having LSR velocities between 6 and 8 \kms. The spatial and velocity structure of the CO line ratios implies that a detailed 3-dimensional radiative transfer model of the cloud would be necessary for a complete interpretation of our spectral data.

It is likely that the observed Class I, flat, and Class II YSOs in the Serpens core formed in a $\sim$1 pc volume between the flat and/or cold centroid positions. The distributions of projected separations and PAs suggest a co-spatial formation site (Figs. \ref{fig:yso_hist_sep}-\ref{fig:yso_hist_pa}). Our measured \tco~ velocity dispersion of \squig  $2.0 \pm 0.6$ \kms~implies that  the Serpens core YSOs could have formed within a small volume and then diffused away to their observed spatial distributions on a timescale of $\sim10^6$~yr.  

The blue- and red-shifted regions in the \tco~velocity first moment (Fig. \ref{fig:serpens_core}) could be the result of a collective action of outflows, such as a preferred protostellar jet orientation. An alternative explanation is a cloud-cloud collision between the blue-shifted gas and foreground gas along the line of sight (Duarte-Cabral et al. 2010; Duarte-Cabral et al. 2011). 

The ``starless cores" region is likely to be the site of further star formation in Serpens.  The detection of 1.1 mm cold dust emission but no CO or \tco~ emission in that region, suggests that CO is largely depleted due to high density and low termperatures in this part of the Serpens cloud.

In a future paper we will employ a grid of statistical equilibrium/radiative transfer models for CO line emission, incorporating CO excitation and photodissociation, to derive the distribution of total hydrogen (H$_{2}$ and HI) column densities and masses of the Serpens core, Serpens G3-G6, and VV Serpentis regions.   This analysis should resolve the disagreement between previous mass estimates noted by Eiroa et al. (2008). 

\section{Acknowledgments}

This material is based upon work supported by the National Science Foundation Graduate Research Fellowship under Grant No. DGE 1106400. This work was supported in part by National Science Foundation grant AST-0708131 to The University of Arizona, and by the NASA Space Grant Program. We thank the anonymous referee for suggestions that strengthened the discussion and conclusion sections in this paper. We thank Steve Stahler, Chris McKee, Yancy Shirley, Neil Evans, and Kevin Hardegree-Ullman for helpful discussions. We also thank Dr. A. R. Kerr of the National Radio Astronomy Observatory for providing the prototype ALMA Band 6 mixers used in this work. We are grateful to Butler Burton and Greg Schwarz for a prompt submission process and storing our data online with The Astrophysical Journal Supplemental
Series.  The Heinrich Hertz Submillimeter Telescope is operated by the Arizona Radio Observatory, which is part of Steward Observatory at The University of Arizona. 

\clearpage

\clearpage

\clearpage

\clearpage

\begin{figure}
\epsscale{0.8}
\plotone{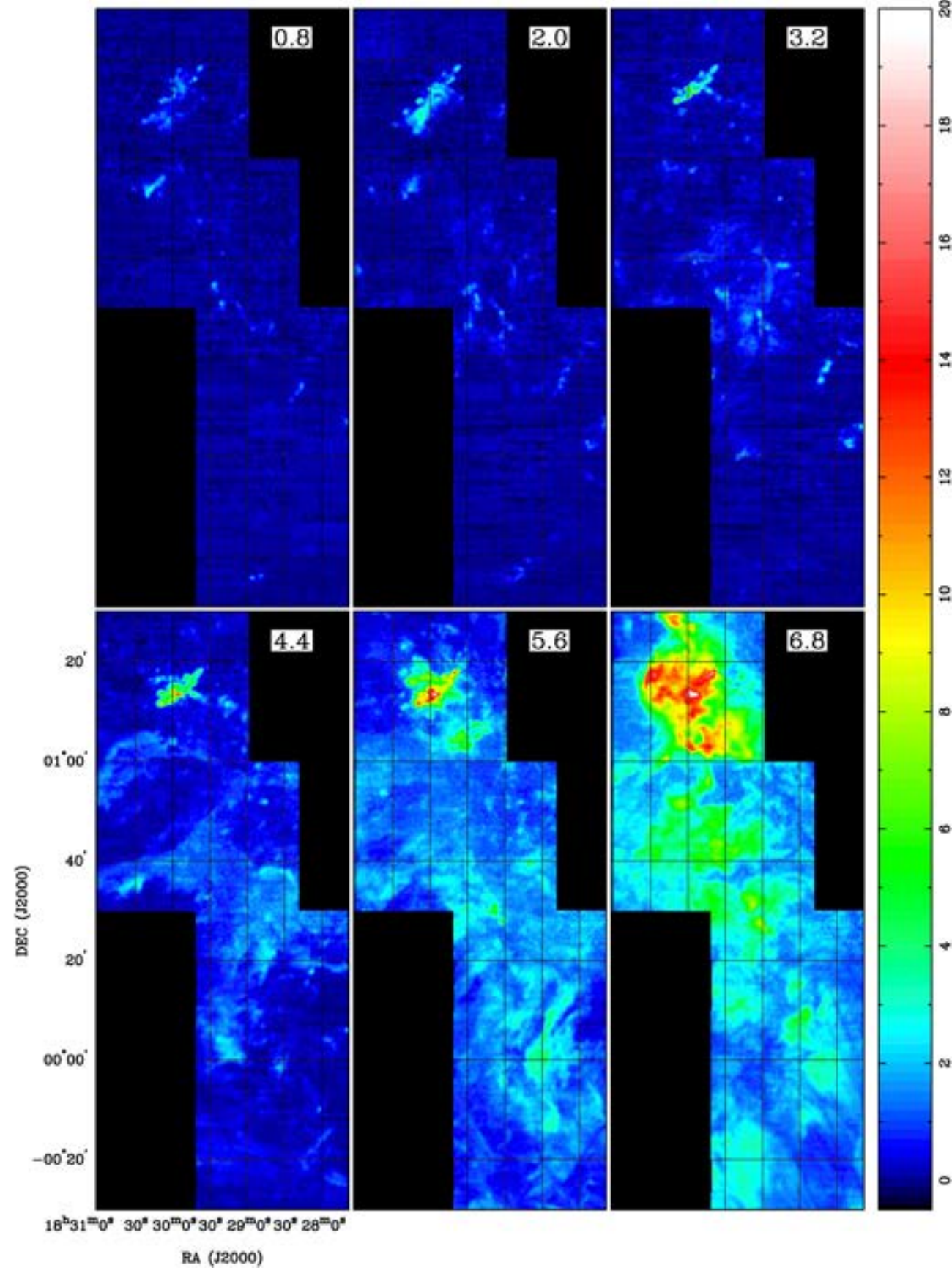}
\caption{CO J=$2-1$ spectral channel maps averaged over 0.6~ \kms~ and spaced 1.2~\kms~ apart.  Mean LSR velocity in upper right.  Color wedge is labeled in main-beam brightness temperature in Kelvins.     
	\label{fig:co_channel}}
\end{figure}
\clearpage

\begin{figure}
\epsscale{0.8}
\figurenum{4} \label{fig4b}
\plotone{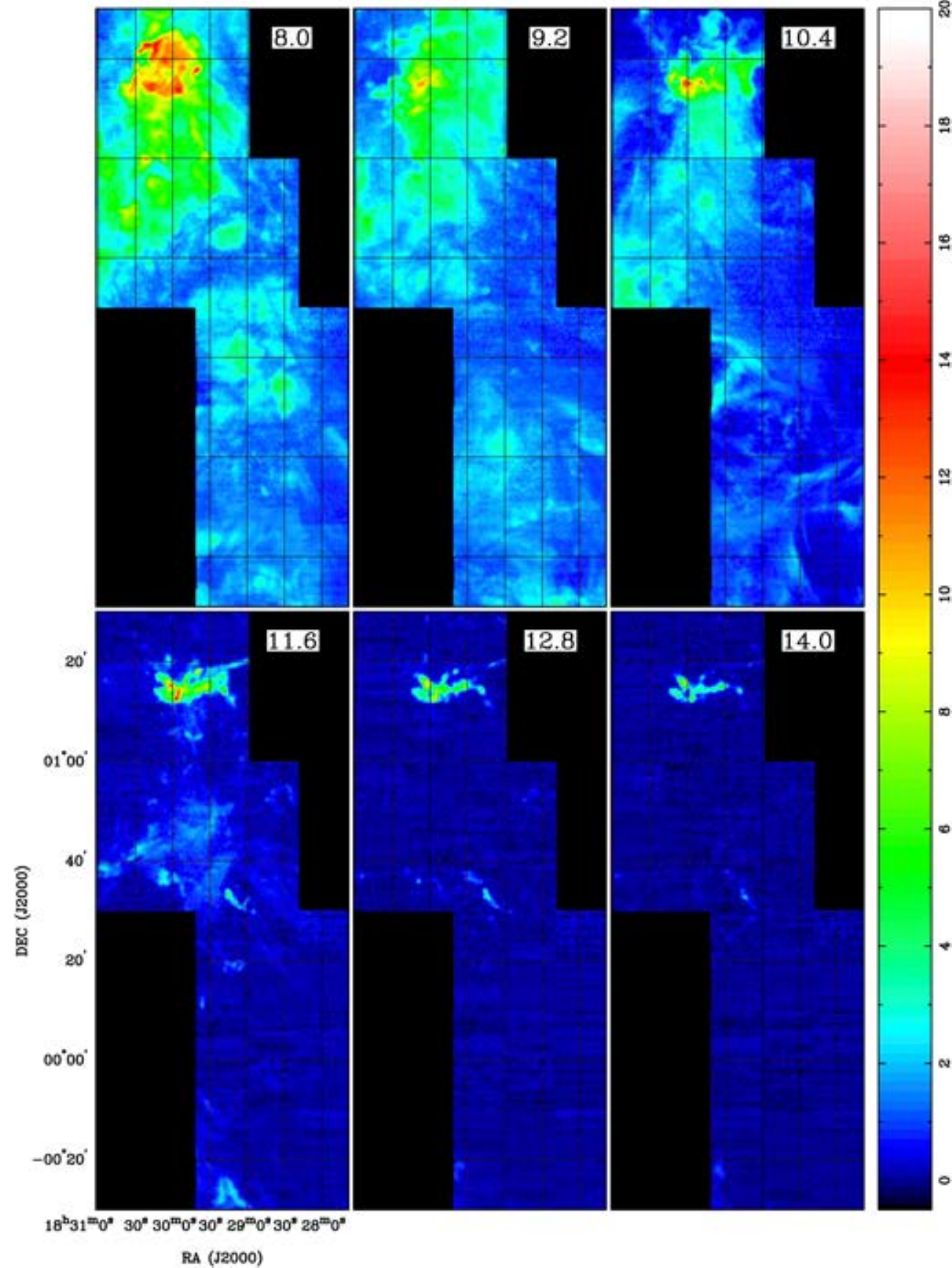}
\caption{ {\it continued} 
	\label{fig:co_channel_2}}
\end{figure}
\clearpage

\begin{figure}
\epsscale{0.8}
\plotone{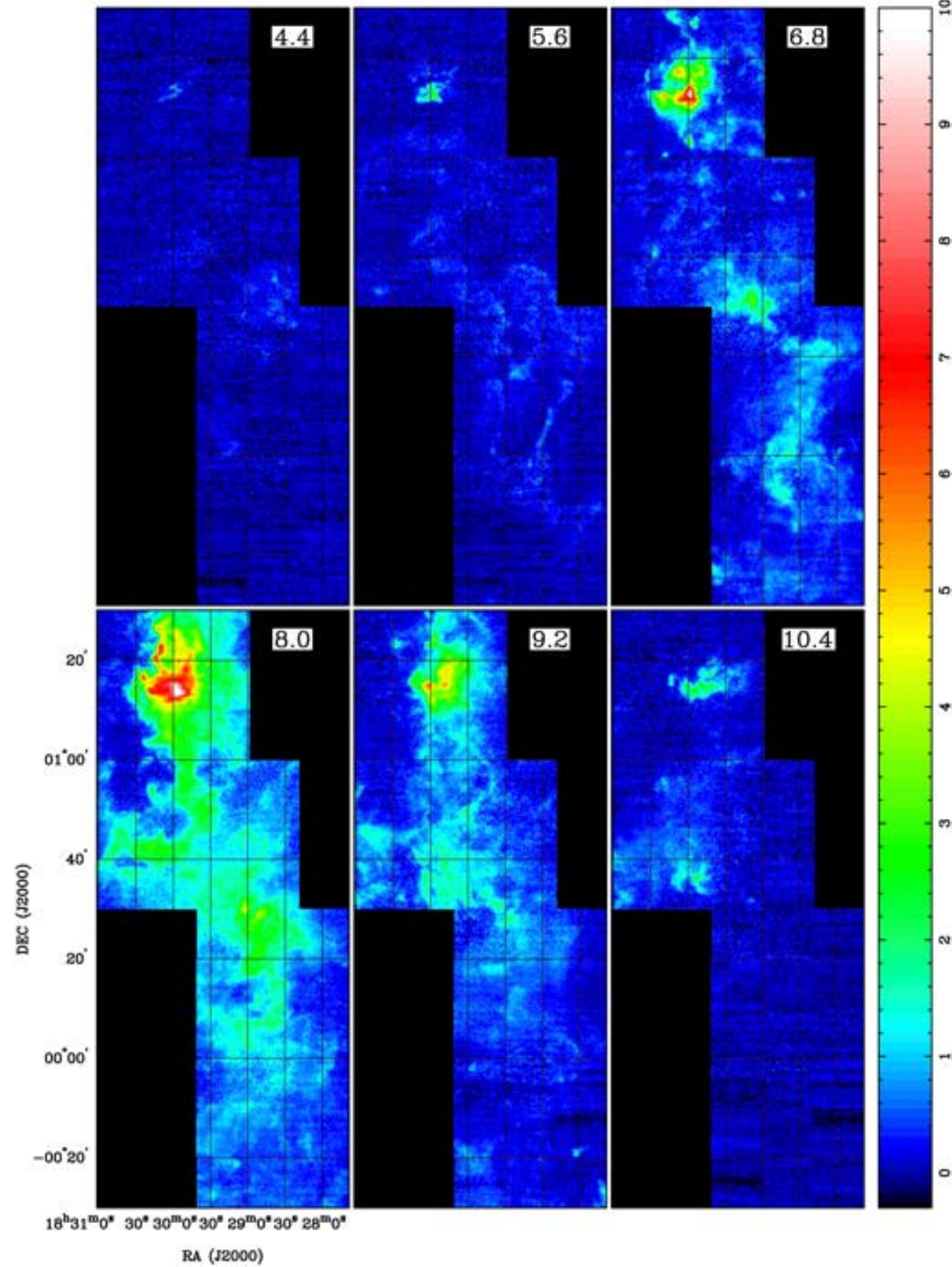}
\caption{\tco~ J=$2-1$ spectral channel maps averaged over 0.6~ \kms~ and spaced 1.2~\kms~ apart.  Mean LSR velocity in upper right.  Color wedge is labeled in main-beam brightness temperature in Kelvins.      
	\label{fig:13co_channel}}
\end{figure}
\clearpage

\begin{figure}
\epsscale{1.2}
\plottwo{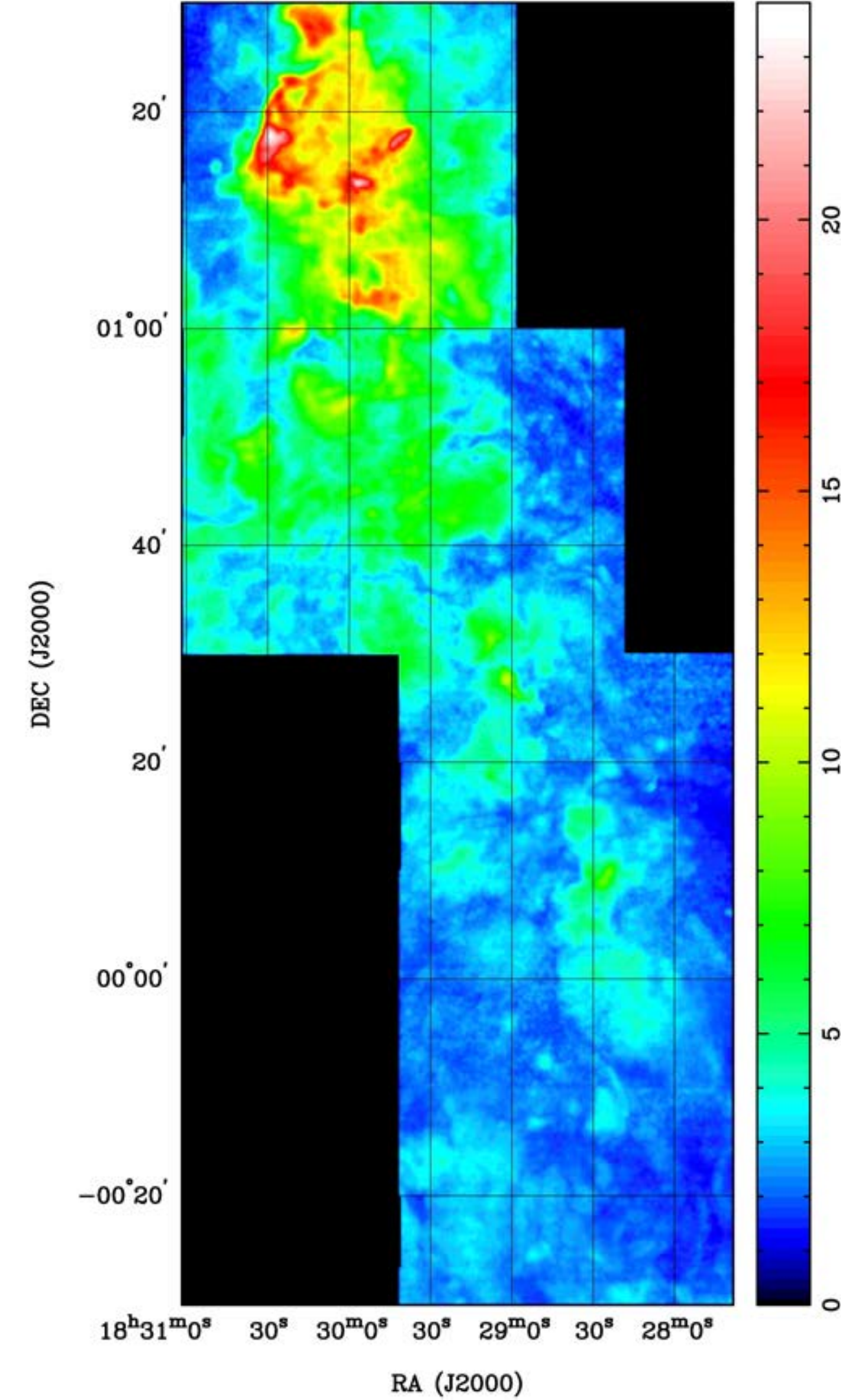}{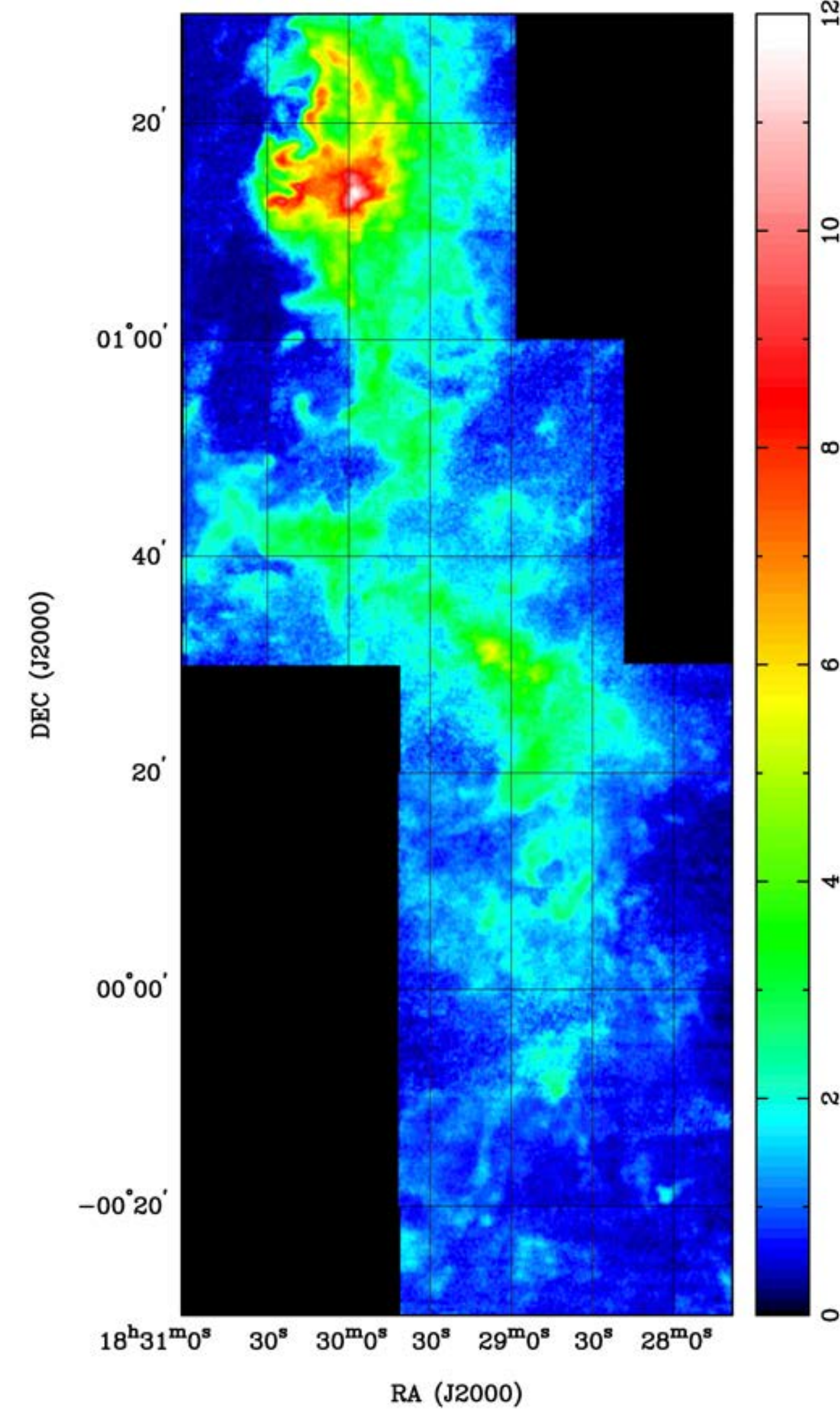}
\caption{Maps of peak brightness temperature, independent of velocity: {\it (left)}, CO J=$2-1$ and {\it (right)}, \tco~J=$2-1$.  Color wedge is in units of brightness temperature (K).  Note that CO emission is detected at all positions in the mapped area.   
	\label{fig:peak_bt}}
\end{figure}
\clearpage

\begin{figure}
\epsscale{1.2}
\plottwo{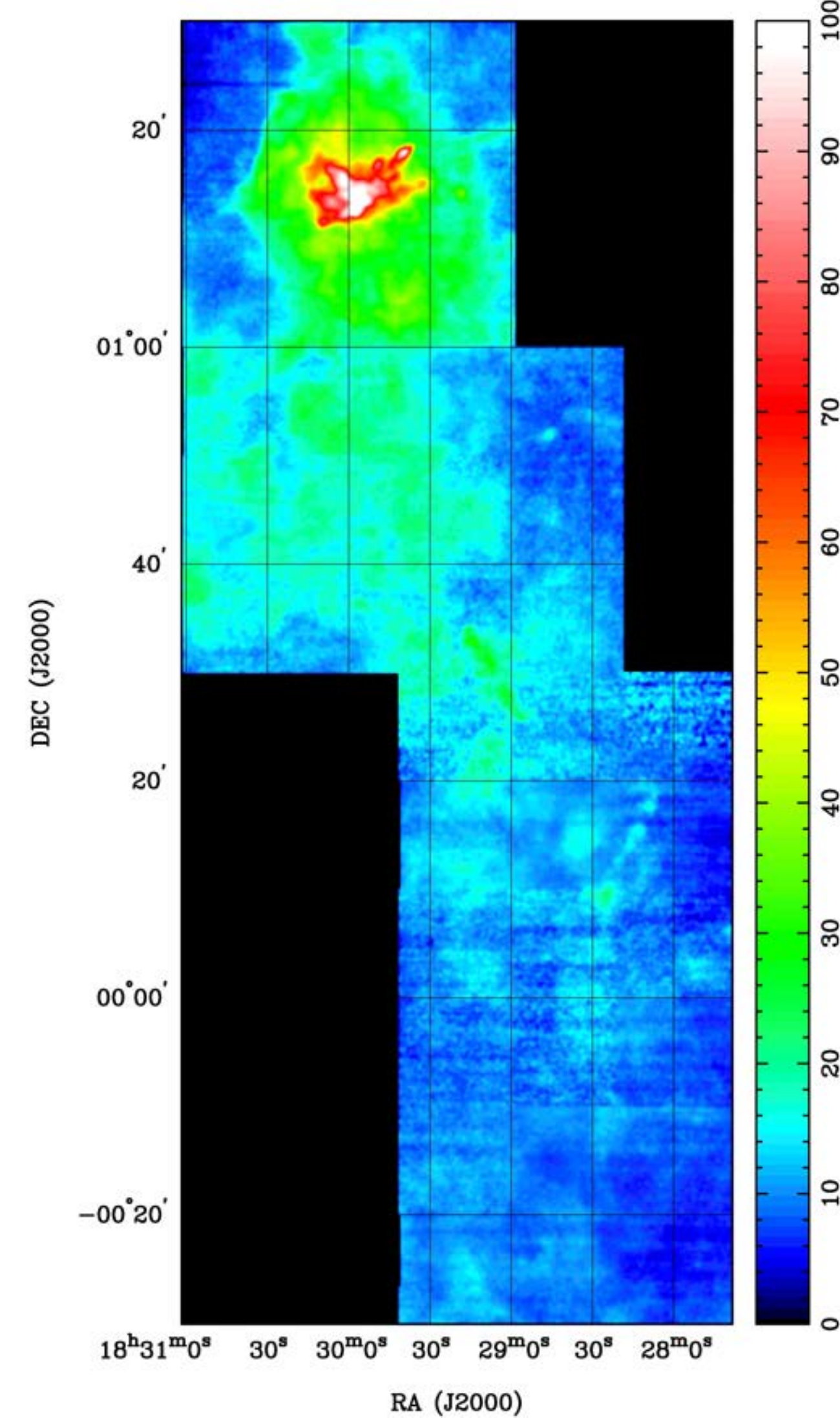}{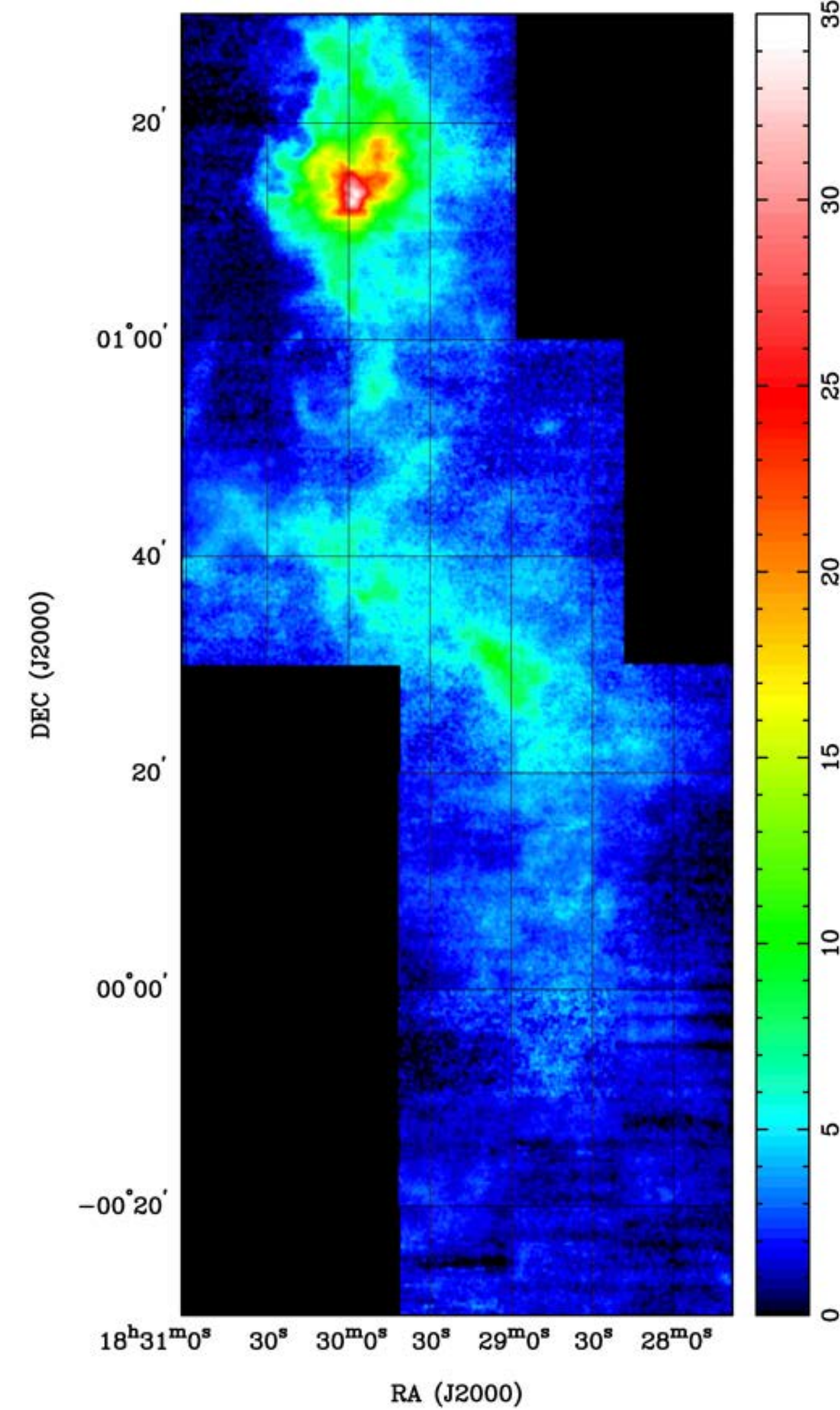}
\caption{Maps of integrated brightness temperature: {\it (left)}, CO J=$2-1$, integrated over velocity range $-1$ to 16 \kms, and {\it (right)}, \tco~J=$2-1$, integrated over velocity range 4 to 12 \kms.  Color wedge is in units of K-\kms.    
	\label{fig:integrated_bt}}
\end{figure}
\clearpage

\begin{figure}
\epsscale{1.2}
\plottwo{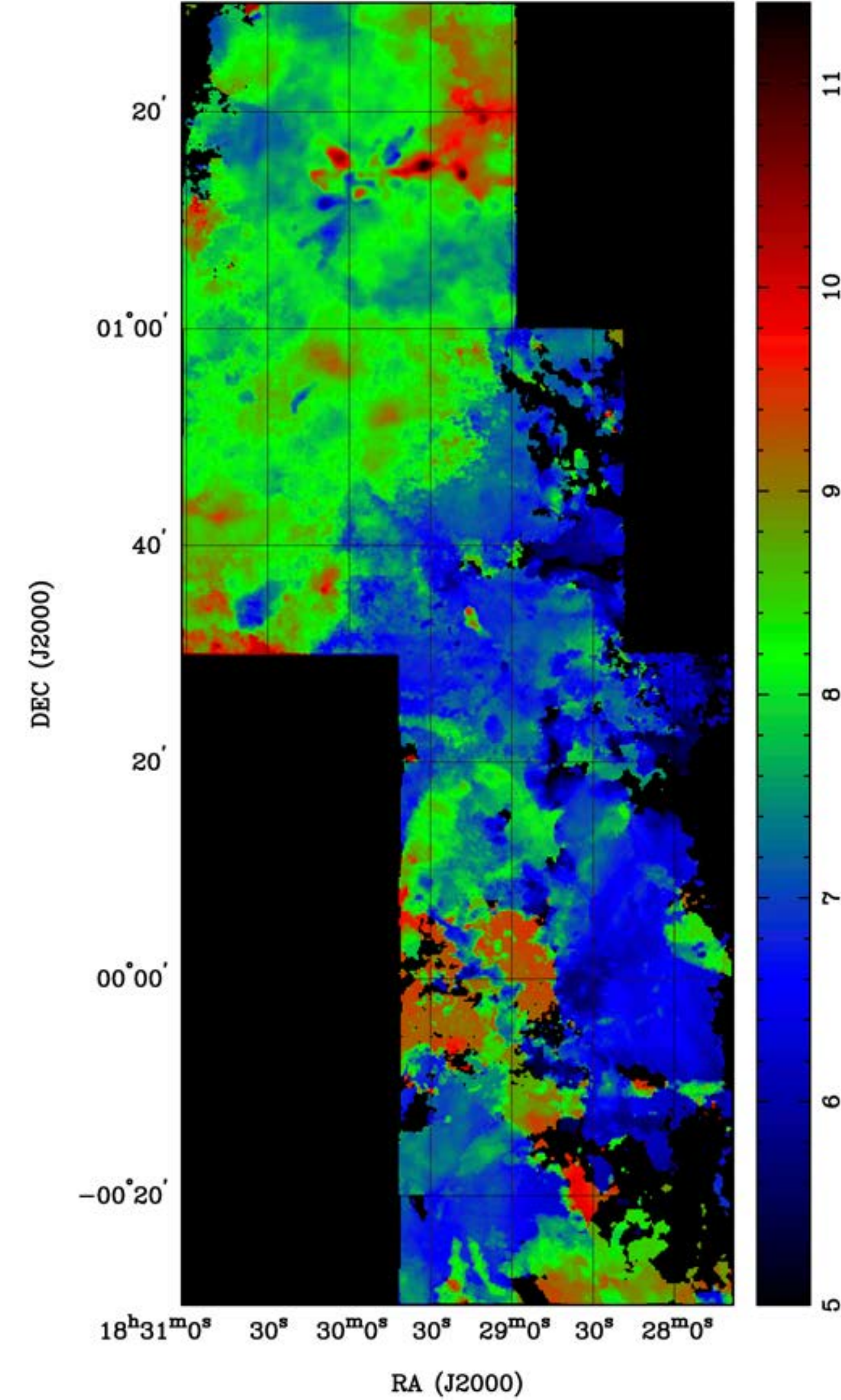}{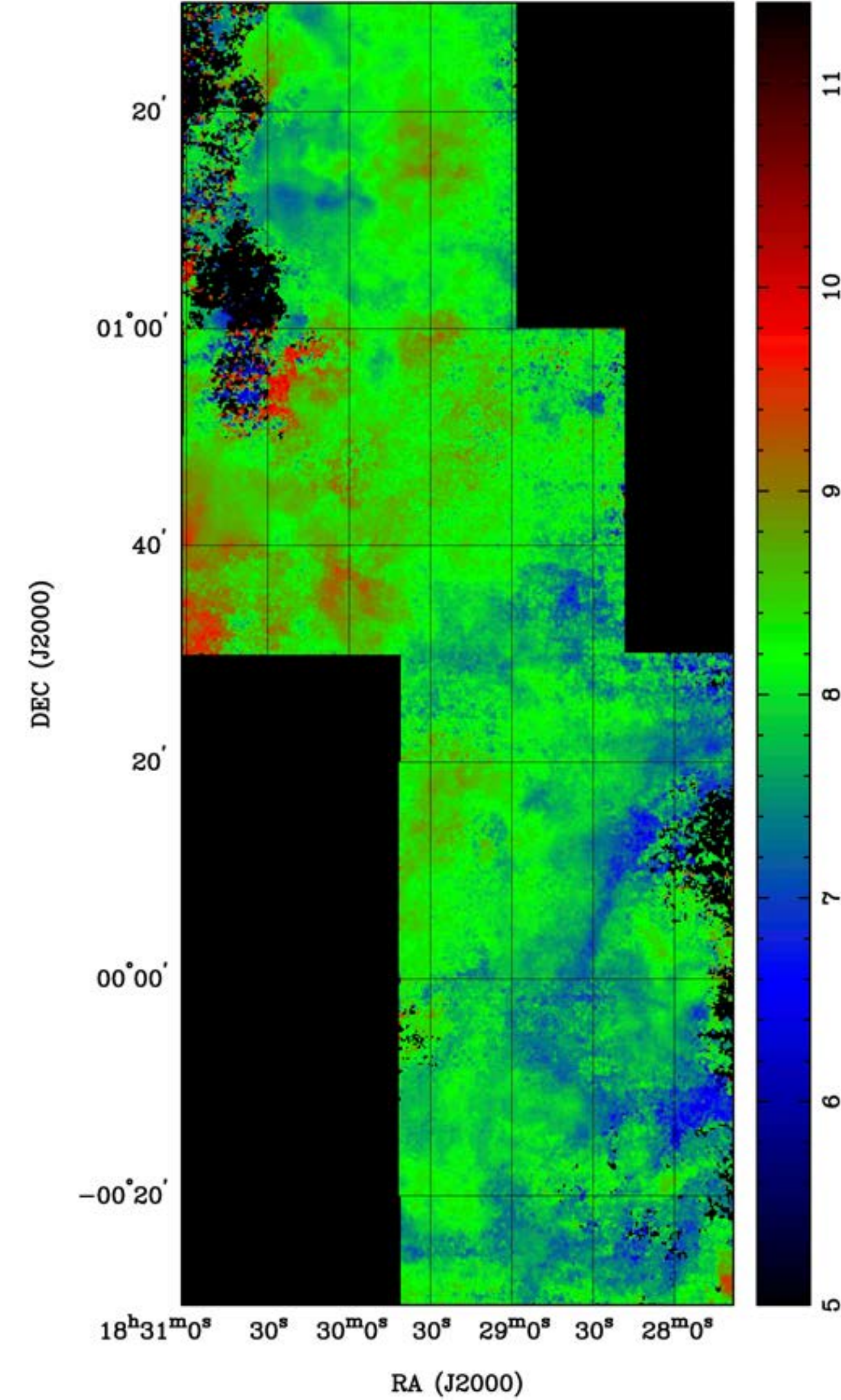}
\caption{Maps of centroid velocity (1st moment of the spectrum): {\it (left)}, CO J=$2-1$ and {\it (right)}, \tco~J=$2-1$.  Color wedge is in units of \kms~ (LSR).  Moments are integrated over the same velocity ranges as in Fig. \ref{fig:integrated_bt}.    
	\label{fig:vel_centroid}}
\end{figure}
\clearpage

\begin{figure}
\epsscale{1.2}
\plottwo{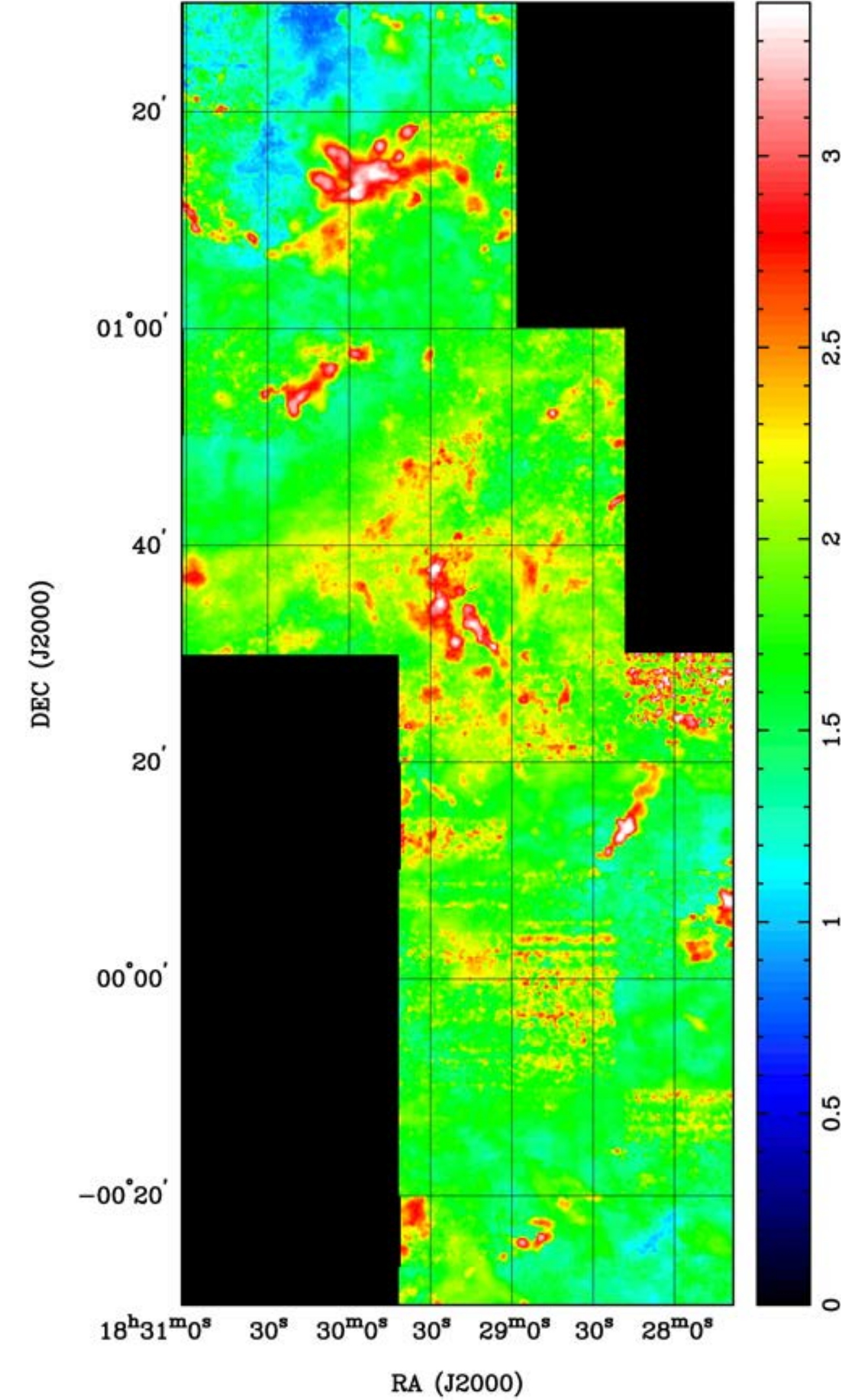}{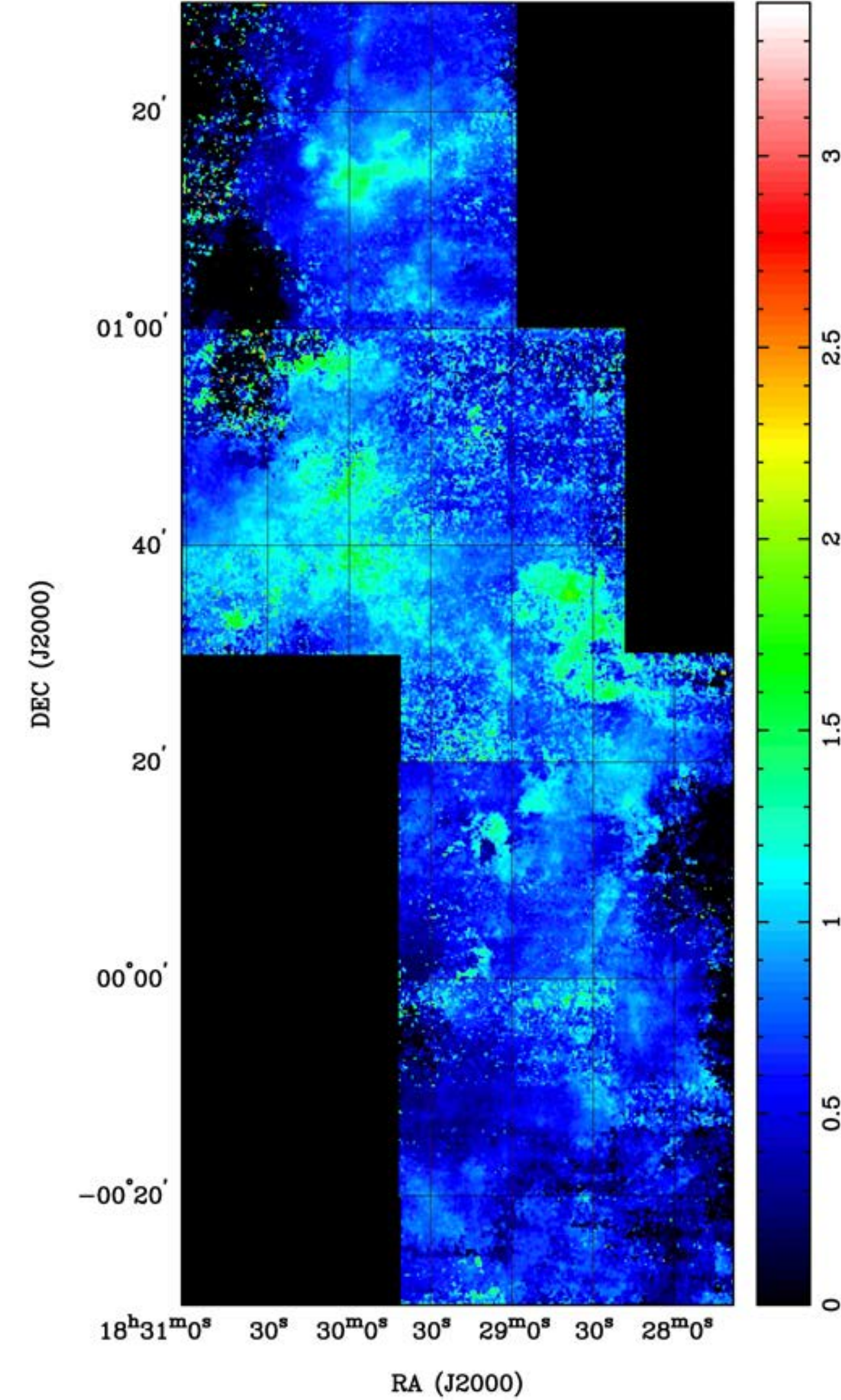}
\caption{Maps of velocity width (2nd moment or dispersion): {\it (left)}, CO J=$2-1$ and {\it (right)}, \tco~J=$2-1$.  Color wedge is in units of \kms.  Moments are integrated over the same velocity ranges as in Fig. \ref{fig:integrated_bt}.  Note that for a Gaussian line shape, the FWHM is the dispersion times a factor of $[8$ ln$(2) ]^{1/2} = 2.355$.    
	\label{fig:vel_width}}
\end{figure}
\clearpage

\begin{figure}
\epsscale{0.8}
\plotone{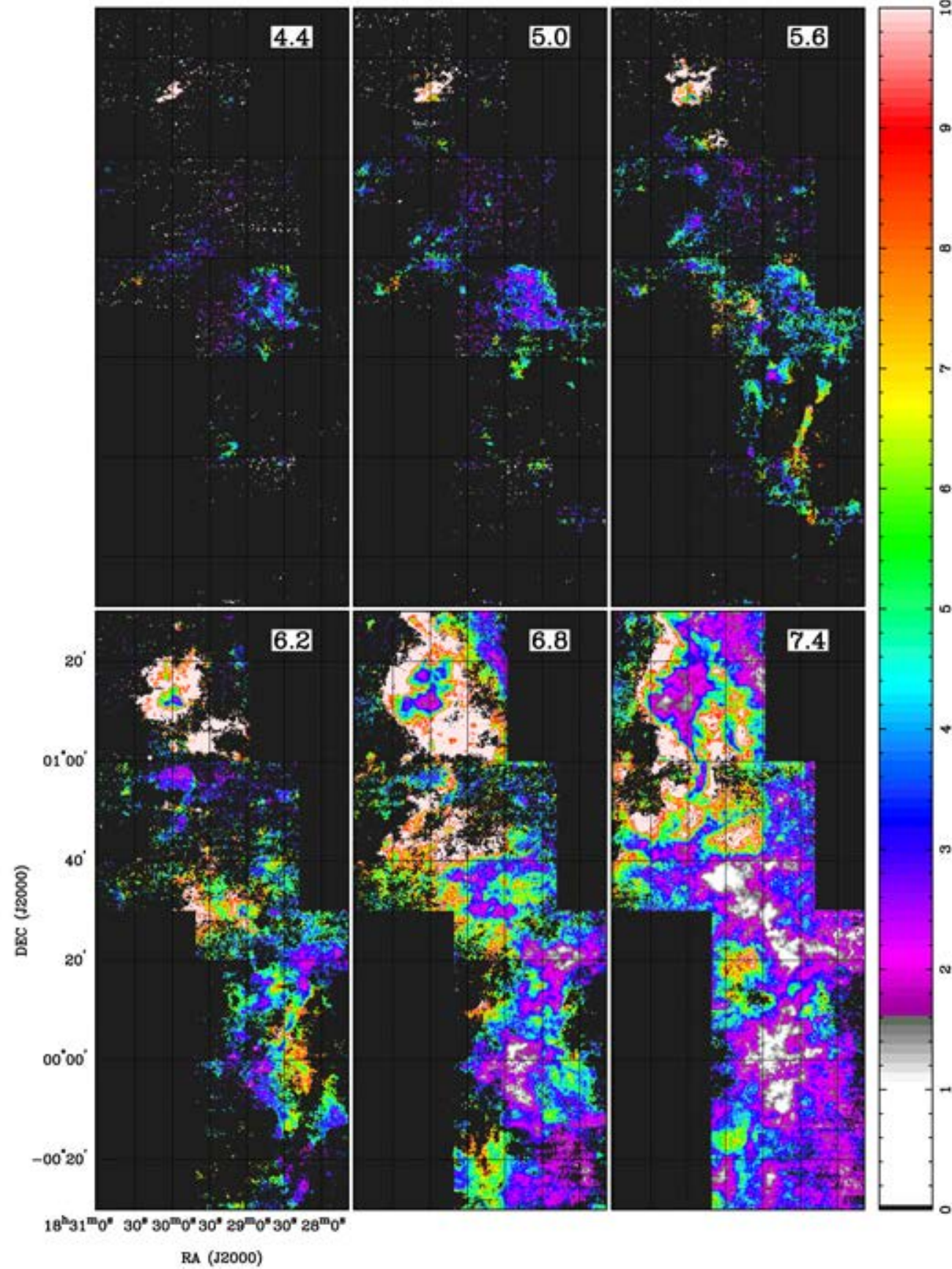}
\caption{Maps of CO/\tco~ J=$2-1$ intensity ratio, averaged over 0.6~ \kms~ and spaced 0.6~\kms~ apart.  Mean LSR velocity in upper right.  Color wedge shows the line ratio $R$ as defined in the text.  Color palette transitions to gray at $R=1.6$, where \tco~ may be optically thick ($\tau = 1$), and to white at $R=1$, indicating that CO is probably self-absorbed.     
	\label{fig:ratio_map}}
\end{figure}
\clearpage

\begin{figure}
\epsscale{0.8}
\figurenum{10}
\plotone{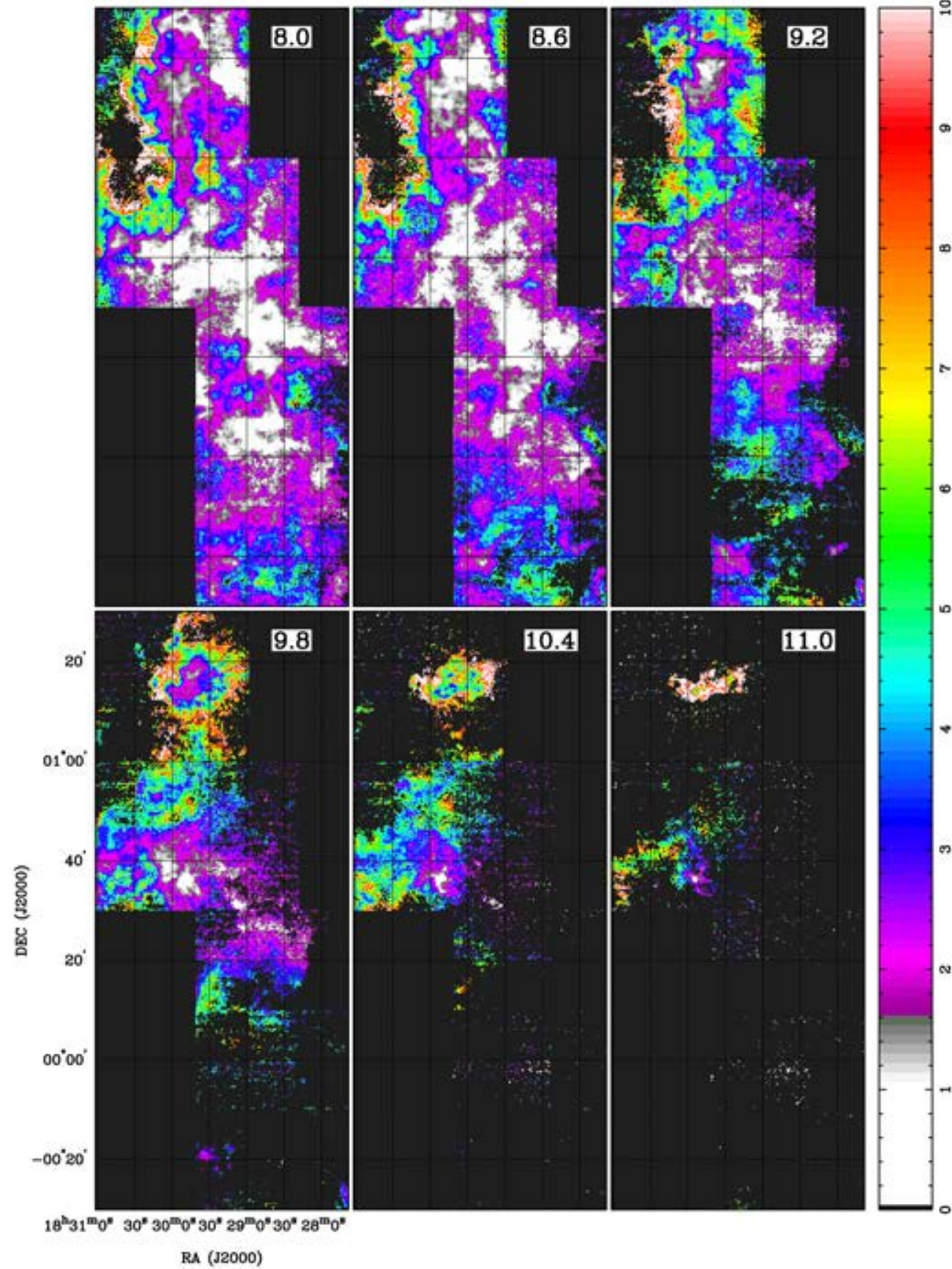}
\caption{{\it continued}   
	\label{fig:ratio_map_2}}
\end{figure}
\clearpage

\begin{figure}
\epsscale{0.7}
\plotone{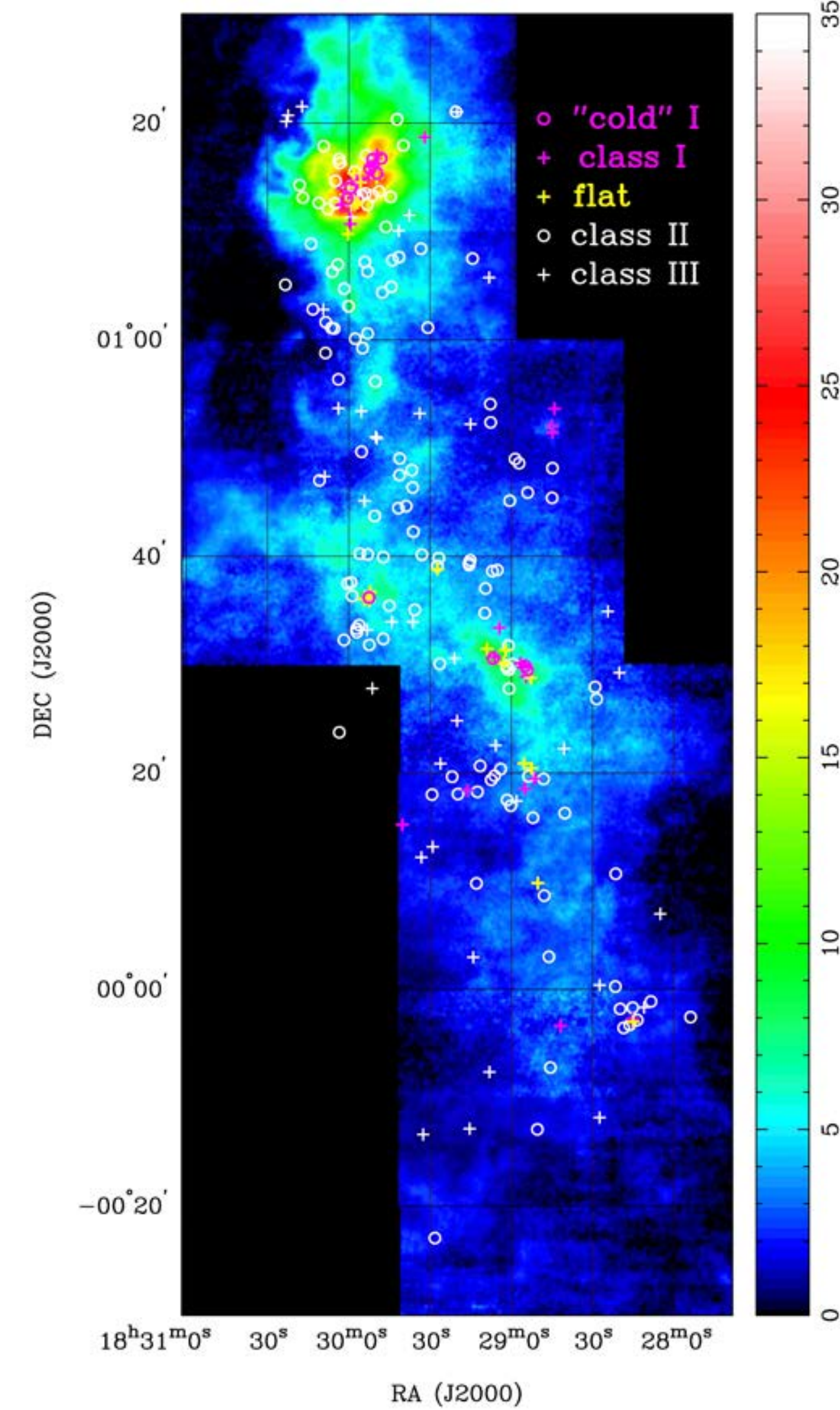}
\caption{Integrated \tco~ intensity map with locations of the 235 Harvey et al. (2007) YSOs marked by $+$'s and $\bigcirc$'s. Symbol key shows SED-classified YSO types. Youngest to oldest: ``cold" Class I (magenta $\bigcirc$'s), all other Class I (magenta $+$'s), Flat (yellow $+$'s), Class II (white $\bigcirc$'s), and Class III (white $+$'s). Color wedge is labelled in K-\kms.    
	 \label{fig:yso_olay}}
\end{figure}
\clearpage

\begin{figure}
\epsscale{1.}
\plotone{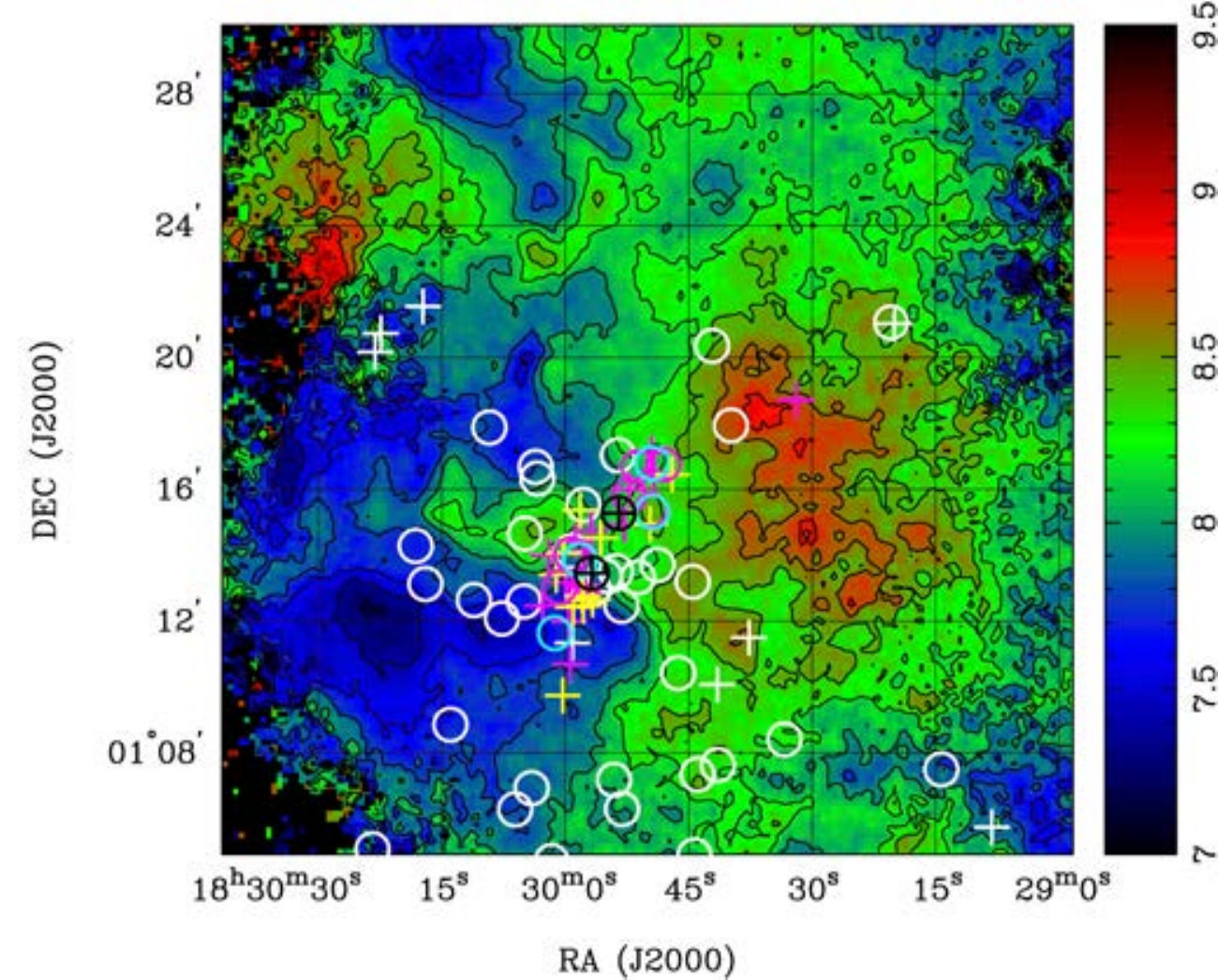}
\caption{Serpens core region map of \tco~ velocity centroid (1st moment) computed over line center (6 to 10.5 \kms) in red-green-blue palette, with positions of YSOs from Harvey et al. (2007) shown as $+$'s and $\bigcirc$'s. The contours are at 7.2, 7.4, 7.6, ..., 8.8 \kms. Symbol key for YSOs is as in Fig. $~\ref{fig:yso_olay}$. Cyan $\bigcirc$'s (diameter 1 arcmin) mark positions of \ntwoh~ J=$1-0$ cores (Testi et al. 2000). Black $\oplus$'s mark the cold (Northern-most $\oplus$)  and flat (Southern-most $\oplus$) centroid positions. These are the unweighted centroid of the R.A., Dec. positions of cold Class I and flat YSOs, respectively.   
	 \label{fig:serpens_core}}
\end{figure}
\clearpage

\begin{figure}
\epsscale{1.}
\plotone{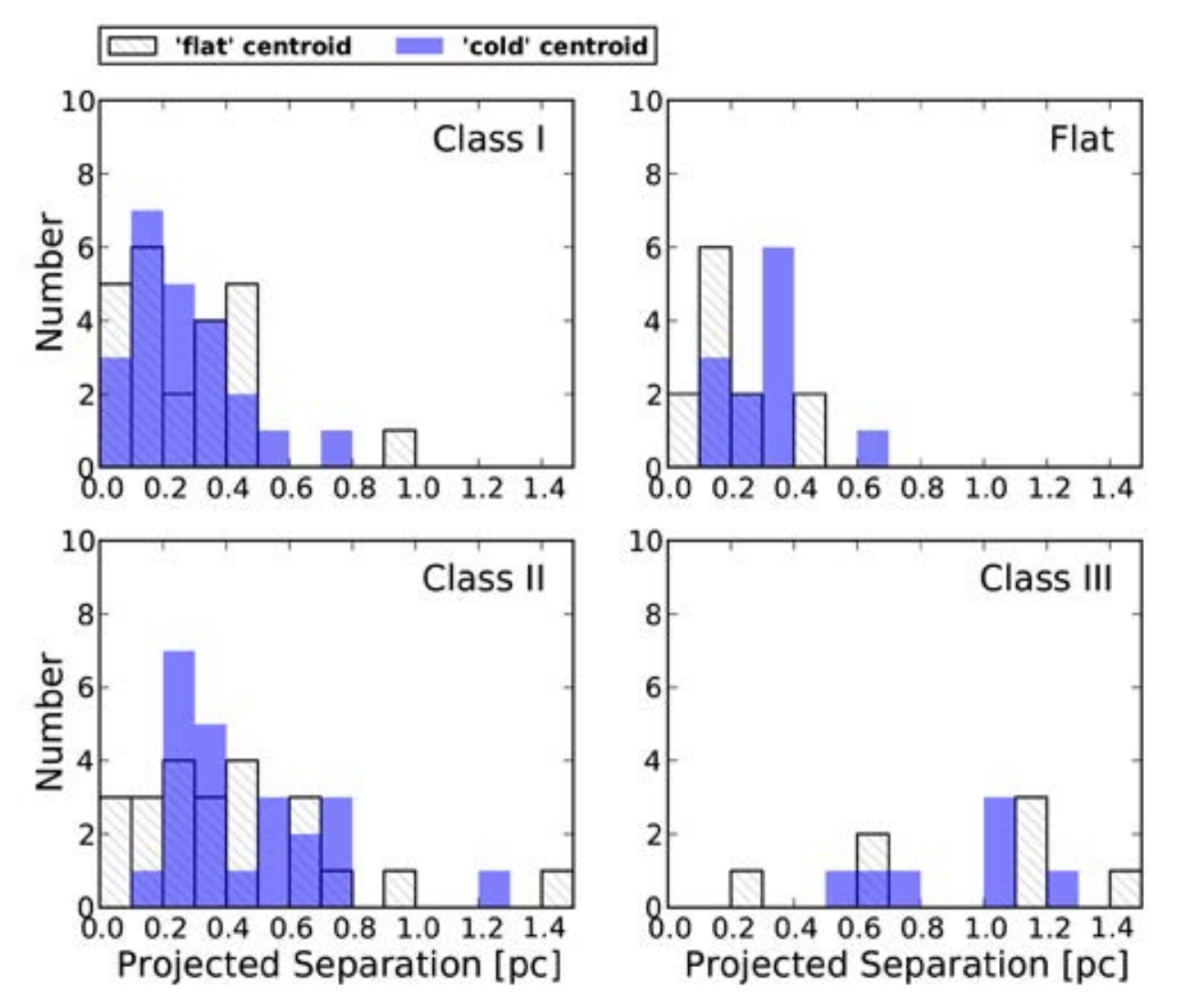}
\caption{Histogram of projected separations between Serpens core YSOs and the cold (blue) and flat (black hatched) centroids. ``Class I" includes cold Class I YSOs.
	\label{fig:yso_hist_sep} }
\end{figure}
\clearpage

\begin{figure}
\epsscale{1.}
\plotone{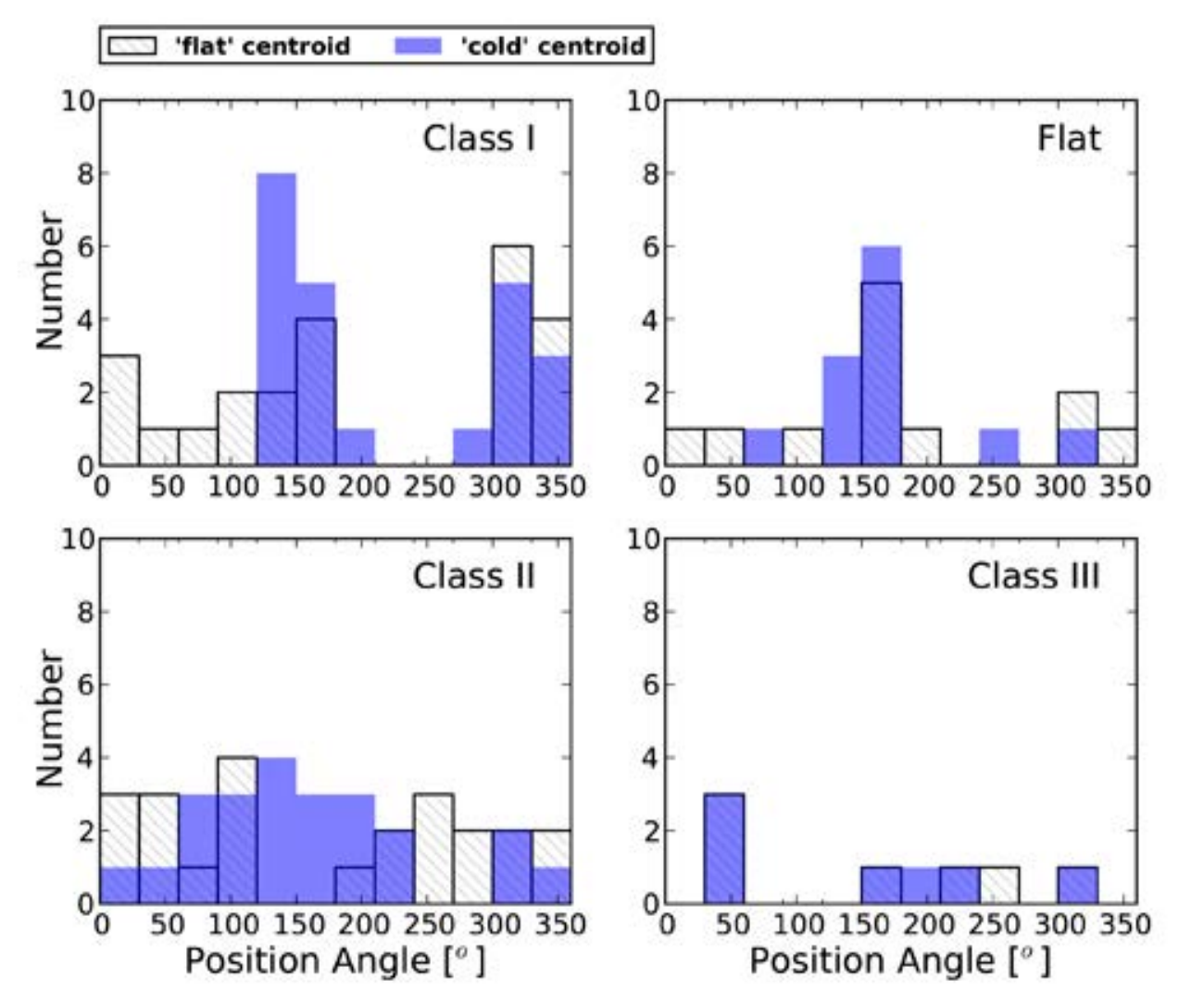}
\caption{Histogram of position angles (PAs) of Serpens core YSOs relative to the cold (blue) and flat (black hatched) centroids. ``Class I" includes cold Class I YSOs. PA = 0\degree~ is North; PA = 90\degree~ is East.  
	\label{fig:yso_hist_pa} }
\end{figure} 
\clearpage

\begin{figure}
\epsscale{0.7}
\plotone{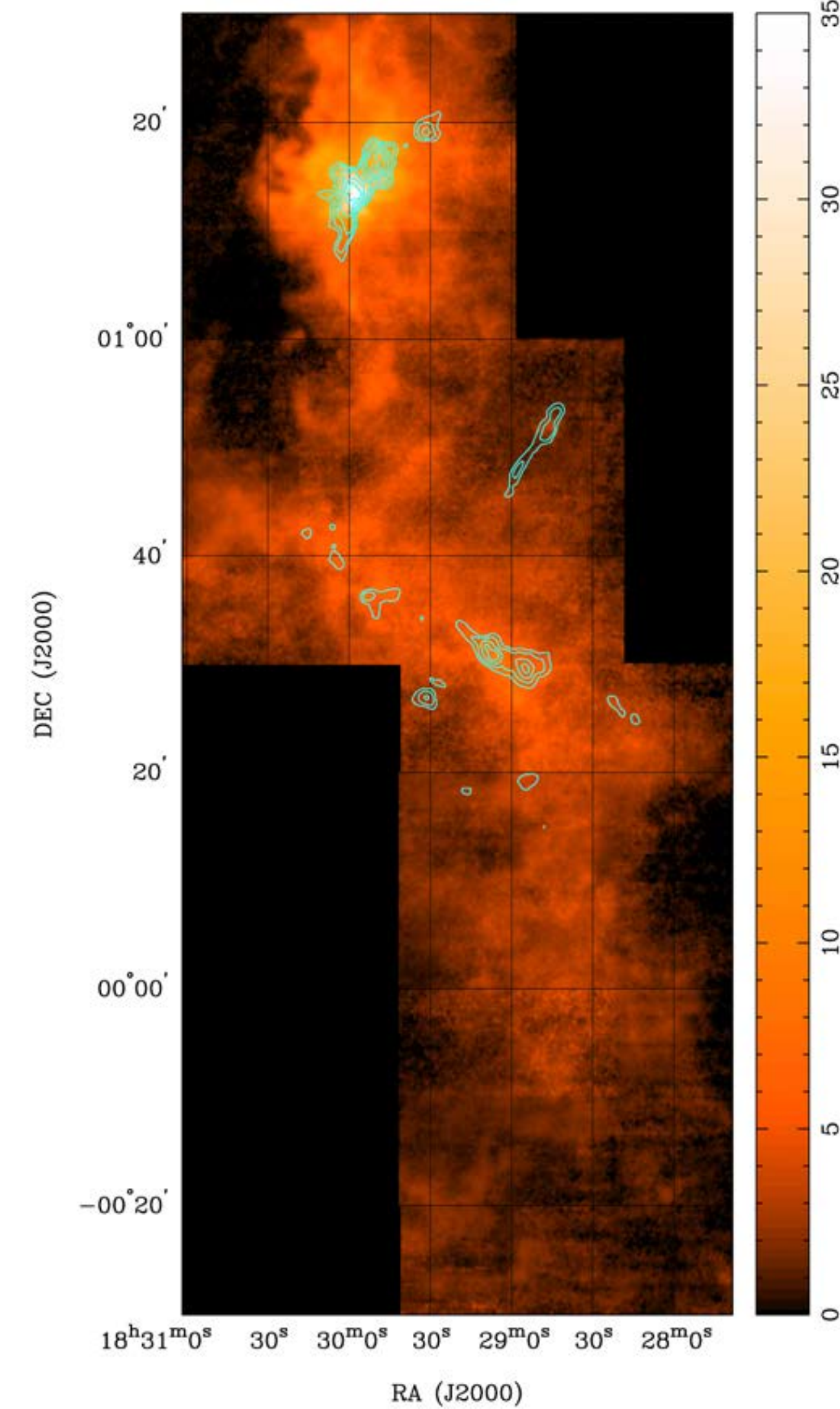}
\caption{Integrated \tco~intensity map in ``heat" colors with contours of Bolocam 1.1 mm continuum emission (Enoch et al. 2007).  Bolocam contour levels are at 5, 10, 20, 40, 80, and 160 times 1$\sigma$ = 11 mJy/beam.  Color wedge is labelled in K-\kms.  The \tco~map has been smoothed to a resolution of 90\arcsec~(FWHM) to match the 1.1 mm image.
	 \label{fig:bolocam_olay} }
\end{figure}
\clearpage

\end{document}